\documentclass{iopjournal}
\usepackage[T1]{fontenc} 
\usepackage[english]{babel}
\usepackage{graphicx}
\usepackage{amsmath}
\usepackage{amsfonts}
\usepackage{float}
\usepackage{ragged2e}
\justifying
\newcommand{\new}[1]{\textcolor{black}{#1}}

\begin{document}
%

\title{Optimal speed-up of multi-step Pontus-Mpemba protocols}

\author{Marco Peluso$^{1,2}$\orcid{0009-0009-3314-9040},  Reinhold Egger$^{2,*}$\orcid{0000-0001-5451-1883}, and Andrea Nava$^{2}$\orcid{0000-0002-6729-3800}}

\affil{$^1$Dipartimento di Scienza Applicata e Tecnologia, Politecnico di Torino, 10129 Torino, Italy}

\affil{$^2$Institut f\"ur Theoretische Physik, Heinrich-Heine-Universit\"at, 40225 D\"usseldorf, Germany}

\affil{$^*$Author to whom any correspondence should be addressed.}

\email{egger@hhu.de}

\begin{abstract}
    \justifying{The classical Mpemba effect is the counterintuitive phenomenon where hotter water freezes faster than colder water due to the breakdown of Newton's law of cooling after a sudden temperature quench. The genuine nonequilibrium post-quench dynamics allows the system to evolve along effective shortcuts absent in the quasi-static regime. 
    When the time needed for preparing the (classical or quantum) system in the hotter initial state is included, we encounter so-called 
    Pontus-Mpemba effects. We here investigate multi-step Pontus-Mpemba protocols for open quantum systems whose dynamics is governed by \new{non-autonomous (\emph{aka} time-inhomogeneous)} Lindblad master equations.  In the limit of infinitely many steps, one arrives at continuous Pontus-Mpemba protocols. 
    We study the crossover between the quasi-static and the sudden-quench regime, showing the presence of dynamically generated shortcuts achieved for time-dependent dissipation rates.
    \new{Considering a two-parameter family of time-dependent rates, the parameters allowing for optimal speed-up conditions are determined.}  Time-dependent rates can also cause non-Markovian behavior, highlighting the existence of rich dynamical regimes accessible beyond the  
    Markovian framework.}
\end{abstract}
 
\section{Introduction}
\label{intro}

Relaxation processes toward equilibrium are commonly described, at a phenomenological level, by Newton's law of cooling \cite{Newton_1701,Maruyama_2021}. In its simplest form, the temperature difference between a system and its environment decays exponentially in time, with a rate determined by the thermal conductance and the heat capacity. This description is valid when the evolution is sufficiently close to equilibrium and the process is quasi-static, meaning that the instantaneous state of the system remains near the equilibrium manifold throughout the evolution. Under these conditions, relaxation times are monotonically increasing functions of the initial distance from equilibrium.

However, out-of-equilibrium dynamics can violate this simple picture. One of the most striking manifestations is the Mpemba effect, originally observed in water freezing experiments \cite{Mpemba_1969}. Under certain conditions, a hotter sample can cool down faster than a colder one, rather than lagging behind it as predicted by Newton's law of cooling. In modern terms, the Mpemba effect corresponds to a non-monotonic dependence of the relaxation time on the parameter variations during a sudden quench. The relaxation speed depends on the geometrical properties of the system state space and on how the initial state projects onto the slow modes of the dynamics. If the initial state has a smaller overlap with the slowest decaying eigenmode, it may reach \new{the steady state} faster despite being initially farther away \cite{Lu_2017,Carollo_2021}.

In recent years, these ideas have been extended beyond the thermal relaxation of classical systems \cite{Lasanta2017,Klich2019,Jesi2019,Torrente2019,Gal_2020,Santos2020,Megias2022,Chetrite2021,Schwarzendahl2022,Walker2023,Ibanez2024,Adalid2024,Vu2025,Bisson2025}. In particular, Mpemba effects have been investigated \new{for closed quantum systems as well as for open quantum systems in contact to external reservoirs, driving them toward equilibrium (thermal) or  nonequilibrium (current-carrying) steady states \cite{Nava2019,Kochsiek2022,Rylands2023,Nava_2024,Murciano_2024,Liu2024,Liu2025,Moroder2024,Ares2025b, Zatsarynna2025,Su2025,Wang2024,Giulio2025,Strachan_2025,Turkeshi2024,Qian2025,Chang_2025,Lacerda2025,Westhoff2025,Li2025_2,Yu2025,Bao2025,Das2025,Chattopadhyay2026}. Importantly,} experimental observations of quantum Mpemba effects have been reported in trapped-ion systems \cite{Joshi2024,Shapira2024,Zhang2025}. In the standard quantum Mpemba protocol, a sudden quench at time, say, $t=0$, takes the system abruptly from one dynamical generator to another, projecting the initial state onto the eigenmodes of the dynamics induced by the final-state attractor. Depending on this projection, nontrivial relaxation behaviors may emerge.

While the standard Mpemba effect compares different initial states for the dynamics induced by the same final-state attractor, a different protocol named ``Pontus-Mpemba'' effect in tribute to an anecdote narrated by Aristotle and related to the region of Pontus in Anatolia \cite{Aristotle}, considers a modified scenario \cite{Nava2025}. Instead of varying the initial condition, one modifies the relaxation protocol itself. In its simplest formulation, one prepares both system copies in the same steady state of an initial generator. For the first system copy, one  performs the direct sudden-quench protocol to switch to the final-state parameters. For the second copy, one first performs a quench toward an auxiliary state (leading to the ``heating up'' of the  state) before a second quench is performed to switch once again to the final-state parameters. \new{We note that a two-step temperature quench  was introduced and discussed in Ref.~\cite{Gal_2020} for thermal baths.
The Pontus-Mpemba protocol therefore includes the time cost for preparing the state in a proper intermediate state \cite{Nava2025} or, in the thermal case, for preheating up or precooling down the initial state \cite{Gal_2020}.}  Under certain conditions, this two-step protocol reduces the total time required to reach a prescribed proximity to the target state in comparison to the direct sudden-quench protocol.  In that case, the Pontus-Mpemba effect is realized.

The essential mechanism differs subtly from the original Mpemba scenario. Rather than exploiting the spectral overlap of initial conditions with the slow modes of a fixed generator, the Pontus-Mpemba protocol reshapes the trajectory in state space. The intermediate dynamics can steer the system toward regions where the subsequent relaxation proceeds faster. In geometric terms, the protocol may shorten the path to the attractor or move the system onto manifolds associated with larger decay rates. Recently, the Pontus-Mpemba effect has been explored in different contexts from imaginary-time dynamics to dynamical phase transitions \cite{Nava_2025_2,Yu_2025,Aditya_2025,Santos_2026,Longhi_2026}.

In this paper, we show how this idea can be further generalized by considering not just a single additional quench but a sequence of parameter variations. In the limit of infinitely many infinitesimal quenches, the protocol becomes continuous, and the generator of the dynamics becomes explicitly time-dependent. This continuous Pontus-Mpemba protocol interpolates between two limiting regimes, namely the quasi-static limit where parameters vary slowly compared to the relaxation rate and the system follows the instantaneous steady state such that no speed-up is expected, and the sudden-quench Mpemba limit, where parameters are changed abruptly. Studying a non-autonomous Lindblad master equation for the dynamics of open quantum systems subject to time-dependent driving and dissipation \cite{Dann_2018,Scopa_2019,Hall_2008,Hall_2014}, we find nontrivial regimes in between these two limits, where a time-dependent modulation of the dissipation rates generates extended regions in parameter space where the relaxation becomes faster than for the sudden-quench protocol. 

In fact, a natural and general framework for describing relaxation processes in open quantum systems is given by  the Lindblad master equation \cite{Lindblad_1976,Breuer2007,Marzano2020} in its non-autonomous form \cite{Dann_2018,Scopa_2019,Hall_2008,Hall_2014}. If the system parameters or the couplings to external environments become time-dependent, the generator of the dynamics becomes explicitly time-dependent as well. Such dynamics describe continuous deformations of the relaxation landscape. The attractor becomes a moving target in state space, and the system may lag behind it depending on the rate of parameter variation. Time dependence also opens the possibility of entering a non-Markovian regime \cite{Piilo_2008,Piilo_2009,Megier_2017,Vega_2017,Becker_2025}. Within the time-local description, non-Markovianity is associated with the breakdown of complete-positive divisibility of the dynamical map. Operationally, this may correspond to temporary backflows of information from the environment to the system. In the non-autonomous Lindblad equation picture, non-Markovian features arise if the decay rates change sign during the evolution. Such behavior signals that the dynamics cannot be interpreted as memoryless at all intermediate times.
The interplay between time-dependent parameters and possible non-Markovian effects provides a rich landscape for engineering relaxation pathways. In this paper, we introduce and study continuous Pontus-Mpemba protocols in order to exploit the freedom offered by a non-autonomous dynamics for reshaping the system trajectories. As we show below, such effects can systematically help to accelerate convergence toward a target state.

The framework described above bears a natural relation to quantum optimal control (QOC) theory \cite{Werschnik_2007,Ansel_2024}. In QOC, one seeks for protocols based on time-dependent Hamiltonians to minimize some cost functional, such as the distance to a target state at a fixed final time or the total time required to reach a prescribed accuracy. The continuous Pontus-Mpemba protocol and the theory of QOC are therefore deeply connected. However, there are important conceptual differences. The traditional QOC theory, eventually adapted to open systems \cite{Koch_2016,He_2024,Duncan_2025,Gautier_2025,Chen_2025}, typically focuses on coherent control through Hamiltonian engineering, whereas the Pontus-Mpemba framework emphasizes the importance of controlled dissipation and relaxation rates. Here, the environment is not merely a source of decoherence but an active resource for state preparation. Thus, the study of continuous Pontus-Mpemba effects sits at the intersection of out-of-equilibrium dynamics, open quantum systems, and QOC. It provides a physically transparent and experimentally accessible route to understanding how a time-dependent modulation of dissipation rates can be harnessed to accelerate relaxation toward a given target state.
\new{Below, we consider a specific class of time-dependent rates and, within this class, determine the parameters allowing for optimal speed-up conditions.}

\new{The present framework is also conceptually related to the theory of shortcuts to adiabaticity (STA), where the goal is to reproduce the outcome of an adiabatic transformation in a finite, and typically short, time by suitably engineering the system dynamics. Both approaches share a common objective, bypassing the limitations of quasi-static evolution by designing time-dependent protocols that steer the system efficiently toward a desired target state. This analogy suggests that tools and ideas developed in the STA framework may provide useful insights for optimizing dissipative state-preparation protocols \cite{Wang_2024,Yin2022,Alipour2020,Duncan_2025,Mahunta_2025}.}

The remainder of this paper is structured as follows. In Sec.~\ref{model}, we introduce the model for an open two-level system studied in this work and 
discuss the non-autonomous Lindblad equation framework for Markovian and non-Markovian dynamics. While we mainly focus on the simplest two-level case, the concepts can be directly generalized to arbitrary Hilbert space dimension. In Sec.~\ref{controller_preparation}, we summarize different out-of-equilibrium protocols and how they can be realized. Next, Sec.~\ref{shortcuts} is dedicated to analyzing numerical results for different parameter regimes, leading to a classification of continuous Pontus-Mpemba effects.  In particular, we demonstrate that a finite quench time interval $\tilde t$, with $\tilde t\to 0$ for the standard Mpemba case and 
$\tilde t\to \infty$ for the quasi-static case, can help to accelerate the relaxation speed toward the target state.  Finally, in Sec.~\ref{conclusions}, we offer concluding remarks.

\section{Model and time-dependent Lindblad master equation}
\label{model}

As simple and experimentally accessible model, we here investigate the dynamics of an open two-level quantum system (a spin-$\frac12$) in an effective magnetic field. The coherent system dynamics   is governed by the Hamiltonian
\begin{equation}
    H(t) = \mathbf{h}(t) \cdot \boldsymbol{\sigma}, 
\end{equation}
with a time-dependent field   $\mathbf{h}(t) = \left( h_x(t), h_y(t), h_z(t) \right)$. The standard Pauli matrices are collected in the vector $\boldsymbol{\sigma} = \left( \sigma_x, \sigma_y, \sigma_z \right)$. To account for dissipation and decoherence induced by an external environment,   the  dynamics of the system density matrix $\rho(t)$ is described by a non-autonomous (\emph{aka} time-inhomogeneous or time-local) Lindblad master equation \cite{Breuer2007} (we put $\hbar=1$ throughout),
\begin{equation}\label{eq:Lindblad_ME}
    \frac{ d \rho(t) }{ d t } = \mathcal{L}[\rho(t)]= - i \left[ H(t), \rho(t) \right] + \sum_{ \lambda} \gamma_\lambda (t) \left( L_\lambda (t) \rho(t) L_\lambda ^\dagger (t) - \frac{1}{2} \left\{ L_\lambda ^\dagger (t) L_\lambda (t), \rho(t) \right\} \right),
\end{equation}
where $\{\cdot,\cdot\}$ is the anticommutator.
The first term on the right-hand-side of Eq.~\eqref{eq:Lindblad_ME} is the Liouvillian and describes the unitary evolution due to the system Hamiltonian. The second term is the Lindbladian and encodes the dissipative non-unitary dynamics in terms of a set of Lindblad jump operators $L_\lambda (t)$ with associated transition rates $\gamma_\lambda(t)$. The superoperator $\mathcal{L}(t)$ is the generator of the system dynamics, i.e., $\rho(t)=\Lambda_{(t,0)}[\rho(0)]$ with 
$\Lambda_{(t,0)}=\mathcal{T}e^{\int_0^t\mathcal{L}(s)ds}$, where $\mathcal{T}$ denotes time ordering.  In general, for a $d$-dimensional system Hilbert space (with $d=2$ in our case), if the jump operators form an orthonormal basis set of $d^2-1$ traceless operators satisfying
\begin{equation}\label{canonica_conditions}
    \mathrm{Tr}[L_\lambda (t)]=0, \quad \mathrm{Tr}[L_\lambda^\dagger (t)L_\mu (t)]=\delta_{\lambda\mu},
\end{equation}
the Lindblad equation assumes its canonical form. Every Lindblad equation of non-canonical form can be reduced to its unique canonical form by a unitary transformation. 

The non-autonomous Lindblad equation \eqref{eq:Lindblad_ME} allows for time-dependent parameters characterizing the unitary and the dissipative time evolution,
allowing also for time-dependent jump operators \cite{Dann_2018,Scopa_2019}. 
If all rates satisfy $\gamma_\lambda(t)\ge 0$ at arbitrary times, the composition law
$\Lambda_{(t_2,t_0)}=\Lambda_{(t_2,t_1)}\Lambda_{(t_1,t_0)}$ holds for arbitrary times $t_2\ge t_1\ge t_0 \ge 0$.
Even for the non-autonomous case described by Eq.~\eqref{eq:Lindblad_ME}, the time evolution is then completely positive and trace preserving (CPT). 
The composition law thus implies that the time evolution can be broken into infinitesimal memoryless steps, and thus represents Markovian dynamics. 
However, since the instantaneous generator ${\cal L}(t)$ changes with time, the system evolves along a time-varying state trajectory  which depends
on the previous history. In this sense, Eq.~\eqref{eq:Lindblad_ME} describes a system which retains memory of the previous time evolution \cite{Hall_2008,Hall_2014}. On the other hand,
if at least one of the rates $\gamma_\lambda(t)$ becomes negative, either at all times or for finite time intervals, Eq.~\eqref{eq:Lindblad_ME} is referred to as 
\emph{pseudo-Lindblad equation}. The above composition law then breaks down, and the time evolution cannot be interpreted as a sequence of memoryless steps anymore. In this case, one may encounter transient backflow of information and/or coherence from the environment into the system. Such phenomena are the hallmark of non-Markovian behavior \cite{Piilo_2008,Piilo_2009,Megier_2017,Vega_2017,Becker_2025}. We note that the breakdown of the composition law does not automatically imply that the total map $\Lambda_{t,0}$ violates the CPT property. The resulting dynamics may still be CPT for all initial states, as expressed by the Fujiwara-Algoet conditions \cite{Fujiwara_1999}. 
Even when complete positivity is violated, however, the map may remain positive, i.e., it preserves the positivity of all density operators of the system, although it would fail to do so when extended to a larger Hilbert space. Alternatively, the map may fail to be positive on the entire state space but remain positive (and therefore physically meaningful) on a restricted subset of initial states. See, e.g., Refs.~\cite{Hall_2008,Hall_2014} for a specification of the general conditions for two-level systems. 
We note that pseudo-Lindblad equations have been applied to study non-Markovian Mpemba effects in Ref.~\cite{Strachan_2025}.

For a two-level system, the state evolution under Eq.~\eqref{eq:Lindblad_ME} can be intuitively understood by a geometric representation using the Bloch vector \cite{Zong_2017}. To that end, one expresses $\rho(t)$ as
\begin{equation}\label{eq:density_matrix}
    \rho \left( t \right) = \frac{1}{2} \left[ \mathbb{I} + \mathbf{r} (t) \cdot \boldsymbol{\sigma} \right],
\end{equation}
with  the $2\times 2$ identity matrix $\mathbb{I}$ and the Bloch vector $\mathbf{r}(t) = \left( r_x(t), r_y(t), r_z(t) \right)$ subject to the condition $|\mathbf{r}(t)|\le 1$. Pure states correspond to the surface of the Bloch unit ball, $|\mathbf{r}|=1$, while mixed states belong to its inner part, $|\mathbf{r}|<1$. The state $|\uparrow \,\rangle_z$ corresponds to the Bloch vector $\mathbf{r} = \left( 0, 0, 1 \right)$, i.e., the north pole of the Bloch ball. Similarly, $|\downarrow \,\rangle_z$ is equivalent to $\mathbf{r} = \left( 0, 0, -1 \right)$, i.e., the south pole.
In this paper, we only study pseudo-Lindblad equations which satisfy positivity for the initial states under consideration. This condition implies that the Bloch vector satisfies
$|\mathbf{r}(t)|\le 1$ for all times \cite{Hall_2008,Hall_2014}. 
 
We next note that different measures for the degree of non-Markovianity for the system dynamics governed by Eq.~\eqref{eq:Lindblad_ME} have been proposed \cite{Rivas_2010,Hall_2014,Breuer_2016}. One such measure is based on the uniqueness of the canonical form of the Lindblad equation. For each dissipation channel, one defines the function
\begin{equation}\label{singlenonM}
    \mathcal{F}_\lambda (t)=-\int_0^t ds \, \mathrm{min}[0, \gamma_\lambda(s)] .
\end{equation}
The total amount of non-Markovianity  then follows by summing Eq.~\eqref{singlenonM} over all dissipation channels \cite{Hall_2014},
\begin{equation}\label{measure-nonmarkovianity}
    \mathcal{F} (t)=\sum_{\lambda=1}^{d^2-1} \mathcal{F}_\lambda (t).
\end{equation}
Note that to compute $\mathcal{F}(t)$, access to the full system dynamics is not 
needed: knowing the time-dependent rates $\gamma_\lambda(t)$ is sufficient.
Clearly, if we always have $\gamma_\lambda(t)\ge 0$, one finds $\mathcal{F}(t)=0$, signalling the absence of non-Markovianity.  
However, one obtains $\mathcal{F}(t)>0$ if at least one of the rates becomes negative in some time interval. Below, we consider these measures for $t\to \infty$. Related experimentally accessible signatures for non-Markovianity are based on the increase of the trace distance \cite{Breuer_2009} or on the dynamics of entanglement quantifiers \cite{Rivas_2010}. However, for at least two dissipative channels, these alternative measures may not reliably witness non-Markovianity anymore. 

We here include three dissipative channels, corresponding to excitation, relaxation, and pure dephasing along the $z$-direction. 
The associated Lindblad jump operators are assumed to be time-independent, but the transition rates $\gamma_\lambda(t)$ are time-dependent. 
Specifically,
\begin{align}
    \left\{ L_\lambda \right\} = \left\{ \sigma_+, \, \sigma_-, \, \sigma_z \right\}, \quad
    \left\{ \gamma_\lambda(t) \right\} = \left\{ \gamma_+(t), \, \gamma_-(t),\, \gamma_z(t) \right\},
\end{align}
with $\sigma_\pm = \left( \sigma_x \pm i\sigma_y\right)/2$. This set of jump operators satisfies the canonical conditions \eqref{canonica_conditions}.
For a two-level system, the Lindblad equation \eqref{eq:Lindblad_ME} is equivalent to an affine linear differential equation for the Bloch vector,
\begin{equation}\label{eq:r_time_derivative}
    \dot{\mathbf{r}} (t) = \boldsymbol{\Lambda}(t) \cdot \mathbf{r} (t) + \mathbf{b}(t),
\end{equation}
with the drift matrix $\boldsymbol{\Lambda}(t)$ and the forcing vector $\mathbf{b}(t)$. In terms of the rates $\gamma_\lambda(t)$ and the field $\mathbf{h}(t)$, we find
\begin{equation}
    \boldsymbol{\Lambda}(t) =2\left(\begin{array}{ccc}
-\frac{\gamma_{+}+\gamma_{-}}{4}-\gamma_{z} & -h_{z} & h_{y}\\
h_{z} & -\frac{\gamma_{+}+\gamma_{-}}{4}-\gamma_{z} & -h_{x}\\
-h_{y} & h_{x} & -\frac{\gamma_{+}+\gamma_{-}}{2}
\end{array}\right), \quad \mathbf{b}= \left(\begin{array}{c}
0\\
0\\
\gamma_{+}-\gamma_{-}
\end{array}\right).
\end{equation}
 In fact, for time-dependent parameters, Eq.~\eqref{eq:r_time_derivative}  admits the closed solution
\begin{equation}\label{eq:indirect_explicit_solution}
    \mathbf{r} (t) = \boldsymbol{\Xi} (t) \cdot \left[ \mathbf{r} (0) + \int_{0}^t ds\, \boldsymbol{\Xi} ^{-1} (s) \cdot \mathbf{b} (s) \right],
\end{equation}
where $\mathbf{r}(0)$ encodes the initial state and the time-dependent 
fundamental matrix is defined as
\begin{equation}
    \boldsymbol{\Xi} (t) = \mathcal{T} \exp \left( \int_{0}^t ds \, \boldsymbol{\Lambda} (s) \right).
\end{equation}
Despite the existence of the analytical solution \eqref{eq:indirect_explicit_solution}, except for few special cases \cite{Tarasov_2021}, it is in general more convenient to numerically solve Eq.~\eqref{eq:r_time_derivative}. 
If all parameters are time-independent, the analytical solution \eqref{eq:indirect_explicit_solution} reduces to
\begin{equation}\label{eq:r_direct_exact}
    \mathbf{r} (t) = e^{ t \boldsymbol{\Lambda}} \cdot \left[ \mathbf{r} (0) + \int_{0}^t ds \, e^{- s \boldsymbol{\Lambda}} \cdot \mathbf{b} \right],
\end{equation}
and the steady state (\emph{aka} attractor) reached at $t\rightarrow \infty$ follows as
\begin{equation}\label{eq:steady-state}
    \mathbf{r}_{\rm ss} = -\boldsymbol{\Lambda}^{-1} \cdot \mathbf{b},
\end{equation}
assuming that $\boldsymbol{\Lambda}$ is invertible. \new{In general, the steady state is not necessarily an equilibrium state since detailed balance can be violated.}
The system dynamics for our model follows from the combination of four different effects, originating from $\mathbf{h}$ and the three rates $\gamma_\lambda$. In particular, (i) the  field $\mathbf{h}$ produces a coherent (Larmor) spin precession of the Bloch vector around the $\mathbf{h}$-axis, preserving the purity of the state, (ii)  $\gamma_z$ squeezes the Bloch vector along the $z$-axis by damping the transverse components, and (iii) the rates $\gamma_{\pm}$ cause an irreversible motion of the Bloch vector toward the (north) south pole of the Bloch ball. 

Finally, let us introduce the trace distance $\mathcal{D}_T$ to quantify the Hilbert space distance between two states $\rho_1$ and $\rho_2$, 
\begin{equation}\label{eq:trace_distance}
    \mathcal{D}_T \left( \rho_1, \rho_2 \right) = \frac{1}{2} \operatorname{Tr} | \rho_1 - \rho_2 |,
\end{equation}
where $|A|=\sqrt{A^\dagger A}$. In the Bloch ball representation for $d=2$, this expression can be easily expressed in terms of the corresponding Bloch vectors,
\begin{equation}\label{eq:trace_distance_euclidean}
    \mathcal{D}_T \left( \rho_1, \rho_2 \right) = \frac{1}{2} | \mathbf{r}_1 - \mathbf{r}_2 |,
\end{equation}
where $|\cdots |$ is the ordinary Euclidean norm. 
Under Markovian  Lindblad evolution, including the time-dependent variant
in Eq.~\eqref{eq:Lindblad_ME}, the trace distance between $\rho(t)$ and $\rho_{\rm ss}$ is a monotonically decaying function of time.  However, in the non-Markovian regime, due to the backflow of information between the system and the environment, contractivity
of the trace distance is not guaranteed anymore \cite{Breuer_2009,Luo_2025}.

\section{Continuous Pontus-Mpemba protocol}
\label{controller_preparation}

By engineering coherent and dissipative processes, one can devise effective protocols for quantum state preparation, in particular by shaping the control parameters in time to steer the system toward a desired target state. State preparation can here be understood as guiding the system along a controlled trajectory toward the desired final state. In the following, we consider different protocols for quantum state preparation based on the non-autonomous Lindblad equation \eqref{eq:Lindblad_ME}, taking advantage of the quantum Mpemba effect for open systems to speed up the dynamics toward the target state. 
We first review the standard one-step quantum Mpemba effect and the two-step Pontus-Mpemba effect. We then generalize it to a multi-step version that allows to further reduce relaxation times \cite{Nava2025}, eventually leading to a continuous Pontus-Mpemba protocol.
For clarity, we here focus on the case $d=2$, i.e., an open two-level system, but the generalization to $d>2$ is in principle straightforward.

\subsection{Standard quantum Mpemba protocol}
\label{direct}

The single-quench quantum Mpemba protocol, which we will refer to as ``direct'' dynamics from now on, consists of two stages. First, at time $t=0$, one initializes the system in a ``starting'' (S) state $\rho(0)=\rho_\mathrm{S}$, corresponding to the steady state of the Lindblad equation for given time-independent parameters $\mathbf{h}_\mathrm{S}$ and $\gamma_{\lambda,\mathrm{S}}$, such that the corresponding Bloch vector satisfies
$\boldsymbol{\Lambda}_\mathrm{S} \cdot \mathbf{r}_\mathrm{S} + \mathbf{b}_\mathrm{S}=0.$
Second, at time $t=0^+$, one performs a sudden quench of the parameters to the time-independent values $\mathbf{h}_\mathrm{F}$ and $\gamma_{\lambda,\mathrm{F}}$ characterizing the ``final'' (F) steady state $\rho_{\rm F}=\rho(t\to \infty)$.   Under the post-quench dynamics, the Bloch vector $\mathbf{r}(t)$ relaxes from $\mathbf{r}(0)=\mathbf{r}_{\rm S}$ toward the steady state corresponding to $\mathbf{r}(t\to \infty)=\mathbf{r}_{\rm F}$, where $\boldsymbol{\Lambda}_\mathrm{F} \cdot \mathbf{r}_\mathrm{F} + \mathbf{b}_\mathrm{F}=0.$

The efficiency of the protocol is quantified by monitoring the time-dependent monotonically decreasing trace distance ${\cal D}_T(\rho(t),\rho_{\rm F})=\frac12| \mathbf{r}(t)-\mathbf{r}_{\rm F}|$. 
In order to extract a relaxation time $\tau$ from the time-dependent trace distance, which then allows for comparing different protocols, we impose a finite but small dimensionless cutoff $\epsilon>0$ such that $\mathcal{D}_T \left( \rho(t), \rho_\mathrm{F} \right)<\epsilon$ for all $t>\tau$. This step is needed since the final state is reached only in the limit $t\rightarrow\infty$.
The parameter $\epsilon$ serves as effective cutoff for the trace distance which reflects the finite precision of the numerical solution of Eq.~\eqref{eq:Lindblad_ME}, setting a threshold below which differences between density matrices are indistinguishable for a given numerical accuracy level. On the experimental side,  $\epsilon$ corresponds to the finite resolution in quantum state tomography, which arises from systematic errors in state preparation and measurement and typically scales $\propto N^{-1}$, where $N$ is the number of measured system copies  \cite{Bartkiewicz_2016, Donnell_2016, Kueng_2017, Bittel_2025}. Consequently, trace distances smaller than $\epsilon$ are operationally irrelevant and cannot be used to reliably distinguish states, neither numerically nor experimentally. States within this $\epsilon$-ball are thus regarded as effectively indistinguishable. In this paper, we set $\epsilon = 10^{-4}$.

\subsection{Two-step Pontus-Mpemba protocol}
\label{two-step}

One of the simplest non-autonomous generalization of the standard quantum Mpemba protocol described above is realized by adding a further intermediate (I) stage, where a second quench is performed at some finite time $t_\mathrm{I}$. We refer to this protocol as two-step Pontus-Mpemba effect, a refinement and generalization of the quantum Mpemba effect which explicitly accounts for the preparation cost of the initial states \cite{Nava2025}.

Rather than comparing relaxation times for two arbitrarily chosen initial conditions, the Pontus-Mpemba framework consists of a two-step protocol, where the ``direct'' dynamics is compared with the dynamics of an identical copy of the system. This second copy starts from the same initial state S with density matrix $\rho(0)=\rho_{\rm S}$, but it is then deliberately driven away along a different direction, toward an auxiliary state A.  At some intermediate state I, the system is coupled to the same environment as the first copy and will then relax toward the same final state F. 
Perhaps counterintuitively, such protocols can often lead to a significantly shorter total preparation time for the second system copy than for the first copy which takes the direct route. 
Formally, the Pontus-Mpemba protocol consists of three stages.  (i) At time $t=0$, we initialize both system copies in the state S.
(ii) At time $t=0^+$, the parameters for the first system copy  are suddenly quenched to time-independent values $\mathbf{h}_F$ and $\gamma_{\lambda,{\rm F}}$, driving it directly toward
the final state F.  For the second system copy,
we instead perform a first sudden quench of the  parameters to fixed values $\mathbf{h}_{\mathrm{A}}$ and $\gamma_{\lambda,\mathrm{A}}$ which drive the system toward an ``auxiliary'' (A) state $\rho_{\mathrm{A}}$. The Bloch vector then evolves continuously toward the steady state solution satisfying $\boldsymbol{\Lambda}_{\mathrm{A}} \cdot \mathbf{r}_{\mathrm{A}} + \mathbf{b}_{\mathrm{A}}=0$.
(iii) Before reaching the state A, the second system undergoes a second quench to the final parameters $\mathbf{h}_\mathrm{F}$ and $\gamma_{\lambda,\mathrm{F}}$ at time $t=t_\mathrm{I}$, where the second system copy has reached the intermediate state $\rho_\mathrm{I}$.  The system then evolves toward the F state. For the second copy, this protocol is described by the non-autonomous Lindblad equation \eqref{eq:Lindblad_ME} with the parameters being piecewise constant functions of time,
\begin{align}
\mathbf{h}(t) & =\begin{cases}
\mathbf{h}_{\mathrm{A}} & 0<t\leq t_\mathrm{I}\\
\mathbf{h}_{\mathrm{F}} & t>t_\mathrm{I}
\end{cases}, \quad \gamma_{\lambda}(t) =\begin{cases}
\gamma_{\lambda,\mathrm{A}} & 0<t\leq t_\mathrm{I}\\
\gamma_{\lambda,\mathrm{F}} & t>t_\mathrm{I}
\end{cases}
.
\end{align}
The relaxation time $\tau$ for the second system copy can then be extracted in the same way as for the direct protocol by demanding $\mathcal{D}_T(\rho(t),\rho_{\rm F})<\epsilon$ for all $t>\tau$,  with the small cutoff $\epsilon$.  We note that 
the trace distance $\mathcal{D}_T \left( \rho(t), \rho_\mathrm{F} \right)$ now consists of a first part (for $0<t\le t_\mathrm{I}$), which generally is not a monotonic function of time, and a second part (for $t>t_\mathrm{I}$) describing monotonic relaxation toward F. (The trace distance is monotonic only when computed with respect to the final steady state of the Lindblad equation. Therefore, during the first time interval where the system is exposed to the A attractor, only $\mathcal{D}_T \left( \rho(t), \rho_{\mathrm{A}} \right)$ is monotonic.) 
We now label by $\rho_{\rm{SF}}(t)$ the density matrix trajectory for the first copy, and by $\rho_{\rm{SIF}}(t)$ for the second copy. Clearly, we have $\rho_{\rm{SF}}(0)=\rho_{\rm{SIF}}(0)=\rho_\mathrm{S}$, $\rho_{\rm{SIF}}(t_\mathrm{I})=\rho_\mathrm{I}$, and $\rho_{\rm{SF}}(t\rightarrow\infty)=\rho_{\rm{SIF}}(t\rightarrow\infty)=\rho_\mathrm{F}$. The same relations hold for the corresponding Bloch vectors $\mathbf{r}_{\rm{SF}}$ and $\mathbf{r}_{\rm{SIF}}$. 

As a function of the relative position between $\rho_\mathrm{S}$, $\rho_\mathrm{I}$, $\rho_\mathrm{F}$, and $\rho_{\rm{SF}}(t_\mathrm{I})$, three different types of Pontus-Mpemba effect can be realized.  The ensuing classification is as follows \cite{Nava2025}.
\begin{itemize}
    \item \textit{Weak type-A}. This case is realized if  $\mathcal{D}_T \left( \rho_\mathrm{I}, \rho_\mathrm{F} \right) < \mathcal{D}_T \left( \rho_{\rm{SF}}(t_\mathrm{I}), \rho_\mathrm{F} \right)$. The A attractor thus drives the system closer to F in a faster manner than under the direct protocol. Once the second system copy is switched to the F attractor, it is already closer to F than the first system copy. \new{We note that, in principle, the system may reach the target state exactly at a finite time $t^*$, i.e., $\rho(t^*)=\rho_{\rm F}$, if the time-dependent dynamics steers the trajectory through $\rho_{\rm F}$ while relaxing toward the auxiliary steady state. One should then choose $t_{\rm I}=t^*$, such that $\mathcal{D}_T \left( \rho_{\rm{SF}}(t_\mathrm{I}), \rho_\mathrm{F} \right)=0$, in order to minimize the relaxation time. However, achieving exact finite-time arrival generally requires  fine-tuning of control parameters and is thus not expected to be robust.} 
    \item \textit{Weak type-B}. This situation corresponds to $\mathcal{D}_T \left( \rho_{\rm{SF}}(t_\mathrm{I}), \rho_\mathrm{F} \right) < \mathcal{D}_T \left( \rho_\mathrm{I}, \rho_\mathrm{F} \right) < \mathcal{D}_T \left( \rho_\mathrm{S}, \rho_\mathrm{F} \right)$. Even though at time $t=t_{\rm I}$, the second system copy is further away from the final state than the first system copy, the attractor A allows the system to approach F from a more convenient direction. This type of Pontus-Mpemba effect is characterized by the presence of a crossing point (similar to the one expected under the standard Mpemba effect) between $\mathcal{D}_T \left( \rho_{\rm{SF}}(t), \rho_\mathrm{F} \right)$ and $\mathcal{D}_T \left( \rho_{\rm{SIF}}(t), \rho_\mathrm{F} \right)$ at some crossing time $t_c>t_{\rm I}$. 
    \item \textit{Strong}: The third type of Pontus-Mpemba effect is realized if the attractor A drives the second system copy further away from both states S and F. 
    Nonetheless, even though the overall trajectory is longer, it requires a shorter time for reaching F. This case is defined by the condition $\mathcal{D}_T \left( \rho_\mathrm{S}, \rho_\mathrm{F} \right) < \mathcal{D}_T \left( \rho_\mathrm{I}, \rho_\mathrm{F} \right)$. It is the least intuitive and most difficult type to realize. Similar to the weak type-B effect, also the strong Pontus-Mpemba effect is characterized by a crossing point between the trace distance curves of both system copies.
\end{itemize} 
\begin{figure}
    \centering
    \includegraphics[width=  0.9\textwidth]{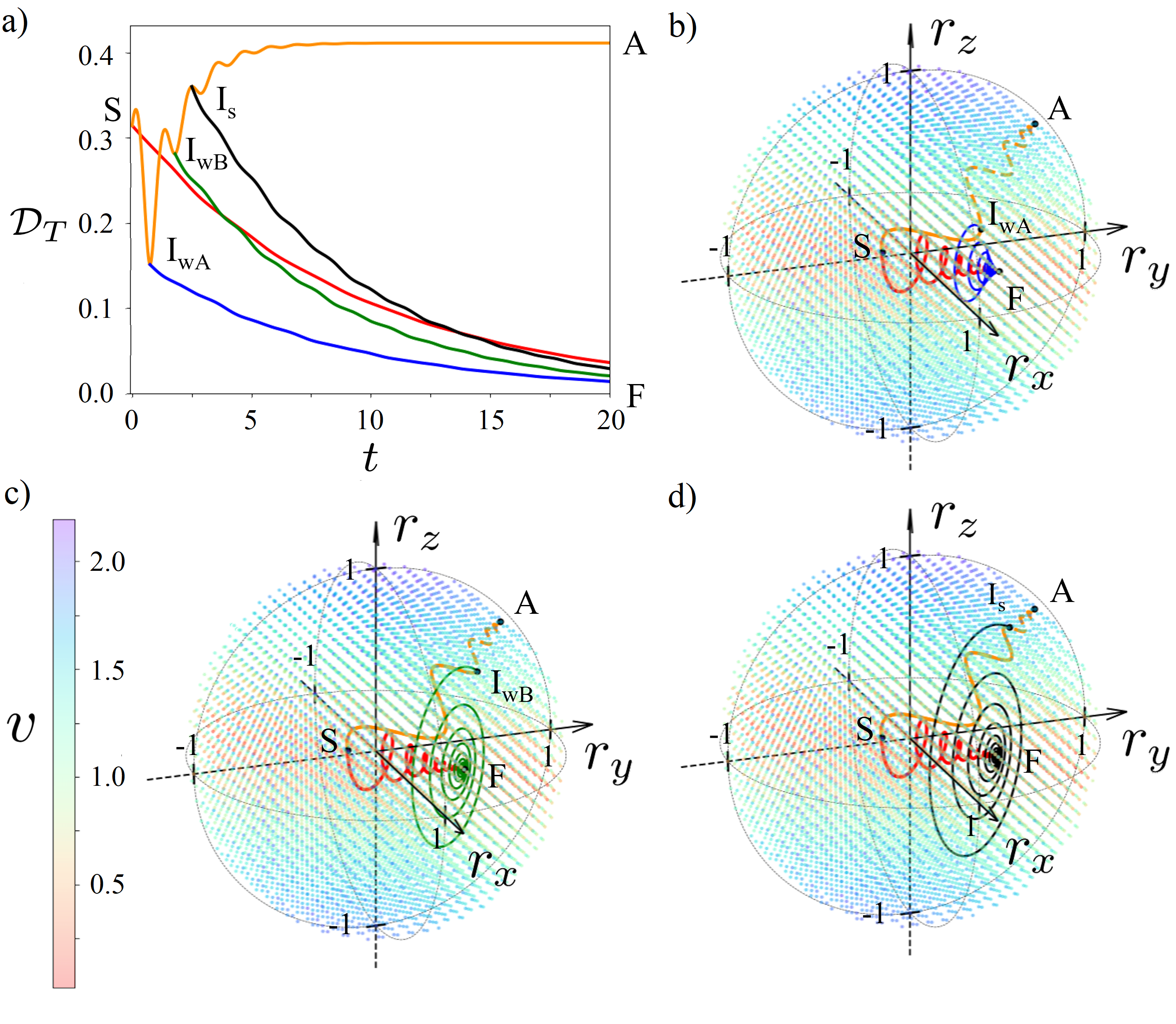}
    \caption{ Two-step Pontus-Mpemba effects in open two-level systems described by Eq.~\eqref{eq:Lindblad_ME}.
    \textbf{(a)}  Trace distance $\mathcal{D}_T \left( \rho(t), \rho_\mathrm{F} \right)$ vs time $t$, comparing the direct  (first system copy) quench protocol to two-step Pontus-Mpemba (second system copy) protocols. We use $|\mathbf{h}_\mathrm{S}|=1$ as energy unit. For both system copies, the initial state S is chosen as steady state, see Eq.~\eqref{eq:steady-state}, for $\mathbf{h}_\mathrm{S}=(0,0.998,0.062)$ and $(\gamma_{+},\gamma_{-},\gamma_{z})_{\rm S}=(0,0.2,0)$, while the final state F is fixed by $\mathbf{h}_\mathrm{F}=(0,-0.966,0.258)$ and  the same rates,  $(\gamma_+,\gamma_-,\gamma_{z})_{\rm F}=(0,0.2,0)$. For the second system copy, the auxiliary state A is chosen as steady state for 
    $\mathbf{h}_\mathrm{A}=(0,2,2)$ and $(\gamma_+,\gamma_-,\gamma_z)_{\rm A}=(1,0,0)$.  The trace distance for the first system copy is shown as red curve. 
    The yellow curve shows the time dependence of  ${\cal D}_T(\rho(t),\rho_{\rm F})$ for the second system copy under the attractor A, where blue, green, and black curves correspond to different choices for the intermediate time $t_{\rm I}$. The weak type-A Pontus-Mpemba effect is realized for $t_{\rm I}=t_\mathrm{I_{wA}}$,
    with the blue curve showing ${\cal D}_T(\rho(t),\rho_{\rm F})$ for $t>t_{\rm I}$. The weak type-B effect with $t_{\rm I}=t_\mathrm{I_{wB}}$ corresponds to the green curve for $t>t_{\rm I}$. Finally, the strong Pontus-Mpemba effect is achieved for $t_{\rm I}=t_\mathrm{I_{s}}$, with the black curve for $t>t_{\rm I}$. All trajectories reach the same final state F for $t\to \infty$.
    \textbf{(b)}  Bloch ball representation of the system dynamics. The red curve represents the direct $\mathrm{S}\rightarrow \mathrm{F}$ trajectory of the Bloch vector, and the yellow curve shows the $\mathrm{S}\rightarrow \mathrm{A}$ trajectory for the second system copy. The blue curve connects $\mathrm{I_{wA}} \rightarrow \mathrm{F}$, corresponding to the blue curve in panel (a), realizing a weak type-A effect. In panels (b)--(d) the velocity field amplitude for the direct protocol is displayed on a color scale using a regular grid within the Bloch ball. The color bar is shown in panel (c).
    \textbf{(c)}  Same as in (b) but for the green trajectory in (a), realizing the weak type-B effect.  \textbf{(d)} Same as in (b) but for the black trajectory in (a), realizing the strong Pontus-Mpemba effect.
    }
    \label{fig:fig1}
\end{figure}
An example illustrating all three  Pontus-Mpemba types in an open two-level system is shown in Fig.~\ref{fig:fig1}, with identically chosen states S, A, and F for all three cases. The  different types follow by choosing different intermediate times $t_{\rm I}$. 
The red curve in Fig.~\ref{fig:fig1} shows the time dependence of the trace distance ${\cal D}_T(\rho(t),\rho_{\rm F})$ for the first system copy (direct protocol).
The trajectory for the second system copy (two-step protocol) instead starts by evolving under the A attractor, see the yellow curve in Fig.~\ref{fig:fig1}(a) for $t<t_\mathrm{I}$, followed by the corresponding blue, green, and black curve for later times $t>t_\mathrm{I}$, with the respective choice for $t_{\rm I}$. If $t_{\rm I}$ is chosen such that the yellow curve stays below the red curve for $t<t_{\rm I}$,  we have the weak type-A Pontus-Mpemba effect, $t_\mathrm{I}=t_\mathrm{I_{wA}}$, where the blue curve shows the dynamics for $t>t_\mathrm{I_{wA}}$.
If the yellow curve is above the red curve at $t_{\rm I}=t_\mathrm{I_{wB}}$, but still below the initial value of the trace distance, we encounter the weak type-B Pontus-Mpemba effect, see the green curve for $t>t_\mathrm{I_{wB}}$. Finally, the black curve represents a strong Pontus-Mpemba effect, $t_\mathrm{I}=t_\mathrm{I_{s}}$, where the yellow curve moves above its initial value. In Figs.~\ref{fig:fig1}(b)--(d),  the Bloch vector trajectories are shown within the Bloch ball for the three types in panel (a).  Note that the two-step trajectory (yellow plus blue, green or black curves) can be longer than the direct trajectory (red) despite of the fact that it leads to faster relaxation toward F.

The Pontus-Mpemba effect can be intuitively visualized with the help of a ``velocity field'' $\mathbf{v}({\bf r},t)$, see Fig.~\ref{fig:fig1}(b)--(d). In the Bloch ball representation of the two-level system, the velocity field follows from the time derivative of the Bloch vector, $\dot{\mathbf{r}}(t)=\mathbf{v}(\mathbf{r}(t),t)$, see Eq.~\eqref{eq:r_time_derivative}. For arbitrary time $t$ and any point $\mathbf{r}$ within the Bloch ball, 
\begin{equation}
    \mathbf{v} ( \mathbf{r}, t ) = \boldsymbol{\Lambda}(t) \cdot \mathbf{r} + \mathbf{b} (t).
\end{equation}
At each point, the velocity field is tangential to the trajectory $\mathbf{r}(t)$ of the Bloch vector evolving under Eq.~\eqref{eq:Lindblad_ME}. It can be generalized to any system dimension $d$ \cite{Nava2025}. Its magnitude $v=|\mathbf{v}|$ allows one to visualize and identify regions in the hyperspace of all possible system states where the state evolves at low vs high speed.
For the first system copy (direct protocol), the velocity field is static and fixed by the state F by means of $\boldsymbol{\Lambda}_\mathrm{F}$ and $\mathbf{b_\mathrm{F}}$. For any initial state S, the Bloch vector $\mathbf{r}(t)$ then evolves along the trajectory $\mathrm{S}\rightarrow \mathrm{F}$. For the second copy (two-step protocol), $\mathbf{r}(t)$ evolves under the velocity field fixed by $\boldsymbol{\Lambda}_\mathrm{A}$ and $\mathbf{b_\mathrm{A}}$ for $t<t_\mathrm{I}$, and under the velocity field of the  F attractor at later times. 
While the direct protocol forces the system to evolve in the velocity field of F, in the two-step protocol, one can wisely choose the auxiliary state A to effectively circumvent low-velocity regions before switching back to the environment corresponding to the final state F. For example, let us consider the cases in Fig.~\ref{fig:fig1}, where the velocity field of the direct protocol is slow around the $r_y$--axis but faster elsewhere. By choosing A such that the system is pushed away from the $r_y$--axis, one can speed up its evolution toward F.
Following such guidelines, it is possible to engineer optimal strategies for reducing the relaxation time, see, e.g., Refs.~\cite{Gal_2020,Santos_2026} for an optimization procedure in the thermal Pontus-Mpemba effect. The two-step Pontus-Mpemba protocol is the simplest speed-up protocol based on the non-autonomous Lindblad equation and provides a rule of thumb for fast state preparation based on an analysis of the velocity field landscape. Indeed, based on the available dissipation channels and for a given target state, one can study the dissipation modes and the corresponding dissipation timescales (which follow from the Lindbladian eigenmodes), which in turn determine the directions (in the space of all possible system states) along which the state dynamics is fast. The auxiliary quench used for the second system copy can then be chosen in order to push the system away from the initial state to more favorable routes.

\subsection{Continuous Pontus-Mpemba protocol}
\label{infnity-step}

The two-step Pontus-Mpemba protocol can be generalized in a natural manner to an $n$-step protocol by performing $n+1$ consecutive quenches,  each driving the system by means of different auxiliary states A$_n$.  Ultimately, the system is again driven to the final state F. In the limit $n\rightarrow\infty$, requiring that the system parameters leading to the attractor A$_n$ form smoothly varying curves in time and approach F for $n\to \infty$, 
one obtains a ``continuous'' Pontus-Mpemba protocol. We here analyze the continuous Pontus-Mpemba effect for systems described by the non-autonomous Lindblad equation  \eqref{eq:Lindblad_ME}, where the sequence $\{{\rm A}_n\}$ is generated by the continuous time-dependent field, $\mathbf{h}(t)$, and dissipation, $\gamma_\lambda(t)$, parameters. For simplicity, we still assume time-independent jump operators. 
We require $\mathbf{h} (0)  = \mathbf{h}_\mathrm{S}$ and $\mathbf{h} (t\rightarrow\infty)  = \mathbf{h}_\mathrm{F}$, as well as
$\gamma_\lambda (0)  = \gamma_{\lambda,\mathrm{S}}$ and $\gamma_{\lambda} (t\rightarrow\infty)  = \gamma_{\lambda,\mathrm{F}}.$  
The instantaneous steady state,
\begin{equation}\label{eq:steady-state-istantaneous}
    \mathbf{r}_{\rm ss}(t) = -\boldsymbol{\Lambda}^{-1}(t) \cdot \mathbf{b}(t),
\end{equation}
then becomes a moving attractor which takes the role of A$_n$ in the $n$-step protocol. If parameters are varied slowly as compared to the system's relaxation rate (which is determined by the spectral gap of the time-dependent Lindblad superoperator), the system approximately follows the instantaneous steady state \eqref{eq:steady-state-istantaneous}. 
This principle enables adiabatic dissipative state preparation, where the final target state is reached by smoothly deforming the steady state of the dynamics \cite{Marr_2003,Wang_2024,Lin_2025}. Clearly, the adiabatic dissipative state preparation imposes an intrinsic lower bound on the total preparation time. Since the system is constrained to follow the instantaneous steady state of the dynamics, the evolution time cannot be shorter than the time required for the steady state (the attractor of the Lindblad dynamics) itself to move from the initial state to the final target state. The adiabatic protocol trades robustness and smoothness for speed: the target state is reached without nonequilibrium transients, but only after a finite deformation time during which the attractor continuously interpolates between the initial and final steady states.
This bound is absent in quench-based or multi-step protocols, where the attractor is displaced instantaneously and the system relaxes toward the target by following an arbitrary nonequilibrium dynamics.

The continuous Pontus-Mpemba protocol allows one to investigate the intermediate regime between the adiabatic  and the sudden-quench (two-step) regime, searching for the optimal characteristic time scale $\tilde{t}$ on which $\mathbf{h}(t)$ and $\gamma_\lambda(t)$ should switch between the values corresponding to S and F. 
The adiabatic (quasi-static) regime is reached for $\tilde{t}\rightarrow\infty$, 
and the sudden-quench protocol is realized for $\tilde{t}\rightarrow 0$.

\section{Dynamical shortcuts}
\label{shortcuts}

We now study the continuous Pontus-Mpemba protocol  by investigating the behavior of the relaxation time $\tau$ (which is obtained by monitoring the trace distance) for a specific time-dependent profile of the dissipation rates $\gamma_\lambda(t)$. 
The results below show that the 
optimal time scale $\tilde t$ is located in between the quasi-static and the sudden-quench regimes.  For the optimal choice of $\tilde t,$  the relaxation time $\tau$ becomes minimal and relaxation to F proceeds at maximum speed. This effect is due to dynamical shortcuts akin to those in standard Mpemba protocols \cite{Lu_2017,Nava_2024}. However, the shortcuts discussed here become available only during the system's time evolution for finite values of $\tilde t$. In particular, they disappear both in the sudden-quench (direct) Mpemba protocol as well as in the quasi-static limit. 

Even for constant jump operators, the general non-autonomous Lindblad equation 
\eqref{eq:Lindblad_ME} allows for an arbitrary time dependence of the Hamiltonian parameters and the dissipation rates. While this generality is useful from a conceptual standpoint, it leads to a high-dimensional control space that is difficult to explore systematically. To reduce the number of free parameters while retaining sufficient flexibility to engineer nontrivial nonequilibrium protocols, we restrict our attention to a specific class of time-dependent rates. In particular, we consider a time-independent Hamiltonian (i.e., a static field  $\mathbf{h}=\mathbf{h}_{\rm S}=\mathbf{h}_F$), while the time-dependent dissipation rates are modelled by an exponential decay together with an oscillatory modulation. For each dissipative channel $\lambda\in\{+,-,z\}$, we assume the same decay rate $\kappa$ and oscillation frequency $\omega$, resulting in 
\begin{equation}\label{eq:gamma_time-dependent}
    \gamma_{\lambda} (t>0) = \gamma_{\lambda, \mathrm{F}} + \left( \gamma_{\lambda, \mathrm{S}} - \gamma_{\lambda, \mathrm{F}} \right) e^{ - \kappa t} \cos{ (\omega t) },
\end{equation}
with the modulation amplitudes $|\gamma_{\lambda, \mathrm{S}} - \gamma_{\lambda, \mathrm{F}}|$. Note that $\kappa$ determines the time scale $\tilde{t} \approx \kappa^{-1}$ discussed above. The parametrization \eqref{eq:gamma_time-dependent} captures several physically relevant features. The exponential envelope ensures that all rates converge to the proper 
final values at long times, thereby guaranteeing the existence of a well-defined
steady state F. The oscillatory component allows for the selective enhancement
or suppression of certain relaxation channels during the early stages of the dynamics.   

The time-dependent rates \eqref{eq:gamma_time-dependent} can assume negative values at intermediate time intervals, thus allowing for non-Markovian behavior. 
Integrating over Eq.~\eqref{eq:gamma_time-dependent}, the measure of non-Markovianity in Eq.~\eqref{singlenonM} is given by
\begin{equation}
    \mathcal{F}_\lambda = {\cal F}_\lambda(t\rightarrow\infty)= - \sum_{ n = 1 } ^ {N_\lambda} \left[ \gamma_{\lambda, \mathrm{F}} t'_\lambda + \dfrac{ \left( \gamma_{\lambda, \mathrm{S}} - \gamma_{\lambda, \mathrm{F}} \right) \omega }{ \kappa ^2 + \omega^2 } \mathit{e} ^{ - \kappa t'_{ \lambda} } \sin{ \left( \omega t'_{ \lambda } \right) } \right]_{ t'_{ \lambda} = t_{ 2 n - 1, \lambda } } ^{ t'_{ \lambda} = t_{ 2 n, \lambda } } ,
\end{equation}
where $t_{2 n - 1, \lambda} $ and $t_{2 n, \lambda }$ (with $ t_{ 2 n - 1, \lambda } < t_{ 2 n, \lambda } $) are solutions of 
\begin{equation}
     e^{ - \kappa t } \cos \left( \omega t\right)  = - \frac{\gamma_{\lambda, \mathrm{F}}}{\gamma_{\lambda, \mathrm{S}} - \gamma_{\lambda, \mathrm{F}} }
\end{equation}
with $n = 1, 2, \ldots,  N_\lambda$ labeling the set of temporal intervals with 
$\gamma_\lambda(t) < 0$. The upper bound is given by 
\begin{equation}
    N_\lambda = \left\lfloor \dfrac{ \omega }{ 2 \pi \kappa } \log \left( \dfrac{ \gamma_{\lambda, \mathrm{S}} - \gamma_{\lambda, \mathrm{F}} }{ \gamma_{\lambda, \mathrm{F}} } \right) - \dfrac{1}{2} \right\rfloor
\end{equation}
with the floor function $\left\lfloor \cdot \right\rfloor$. The onset of the non-Markovian regime for a given dissipation channel is characterized by the conditions $N_\lambda=1$ and $t_{1,\lambda}=t_{2,\lambda}$. This implies the transition curve,
$\omega_\lambda (\kappa) = \kappa/\alpha_\lambda$, which
separates the Markovian from the non-Markovian regime.  Here, $\alpha_\lambda \in \mathbb{R}_{+}$ is the only solution of the implicit equation 
\begin{equation}
    \frac{ e^{ - \alpha_\lambda \left[ \pi - \arctan \left( \alpha_\lambda \right) \right] } }{ \sqrt{ 1 + \alpha_\lambda ^ 2 } } = \frac{ \gamma_{\lambda, \mathrm{S}} - \gamma_{\lambda, \mathrm{F}} }{ \gamma_{\lambda, \mathrm{F}} }.
\end{equation}  
Consequently, non-Markovian behavior with $\mathcal{F} = \sum_\lambda \mathcal{F}_\lambda > 0$ occurs for 
\begin{equation}\label{nMB}
    \omega > \min_\lambda \left\{ \kappa/\alpha_\lambda \right\}.
\end{equation}
In particular, for $\omega=0$, only Markovian behavior is possible.

\begin{figure}
    \centering
    \includegraphics[width=0.9 \textwidth]{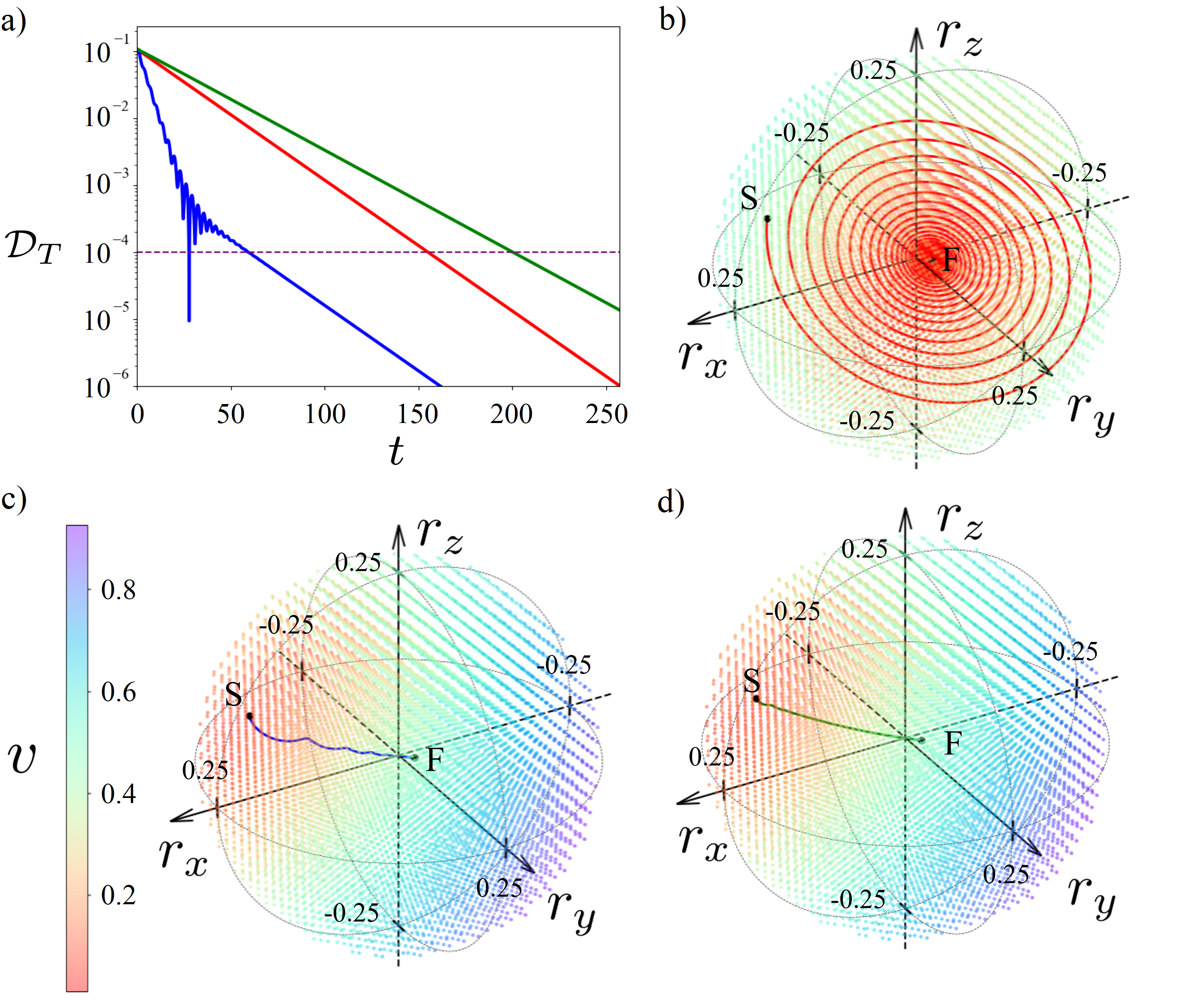}
    \caption{Continuous Pontus-Mpemba protocol for an open two-level system described by Eq.~\eqref{eq:Lindblad_ME} with the time-dependent rates \eqref{eq:gamma_time-dependent}.
    \textbf{(a)}  Trace distance $\mathcal{D}_T \left( \rho(t), \rho_\mathrm{F} \right)$ vs time $t$ for the direct protocol and for the continuous Pontus-Mpemba protocol. We use $|\mathbf{h}_\mathrm{S}|=1$ as energy unit.
    In both cases, the initial state S is the steady state \eqref{eq:steady-state} corresponding to $\mathbf{h}_\mathrm{S}=(0.707,0.707,0)$ and
    $(\gamma_+,\gamma_-,\gamma_z)_{\rm S}=(0.5,0.1,0)$, while the final state F 
    is determined by $\mathbf{h}_\mathrm{F}=\mathbf{h}_\mathrm{S}$ and
    $(\gamma_+,\gamma_-,\gamma_z)_{\rm F}=(0.01,0.05,0)$.
     The trace distance for the direct protocol ${\rm S}\rightarrow {\rm F}$ is shown as red curve.      The green (blue) curve corresponds to the trace distance between S and F for a trajectory described by the time-dependent parameters in Eq.~\eqref{eq:gamma_time-dependent} with $\omega=0$ and $\kappa=0.035$ ($\kappa=0.2$). In all cases, the state F is reached for $t\to \infty$, and the system dynamics is Markovian (since $\omega=0$). The dashed line indicates the cutoff $\epsilon=10^{-4}$ for the trace distance.
    \textbf{(b)}  Bloch ball representation of the system dynamics. The red curve represents the direct sudden-quench curve $\mathrm{S}\rightarrow \mathrm{F}$ for $\mathbf{r}(t)$, corresponding to the red curve in (a). Only the portion of the Bloch ball with $|\mathbf{r}|<0.25$ is shown. The velocity field amplitude for the F attractor is shown using color scales on a regular grid inside the Bloch ball; the color bar is shown in (c).
    \textbf{(c)}  Same as (b) but for the blue curve in (a), realizing the continuous Pontus-Mpemba effect. The shown velocity field corresponds to the attractor S for the initial state.
    \textbf{(d)}  Same as (c) but for the green curve in (a). Here the continuous Pontus-Mpemba effect does not take place. }
    \label{fig:fig2}
\end{figure}

\begin{figure}
    \centering
    \includegraphics[width= 0.9 \textwidth]{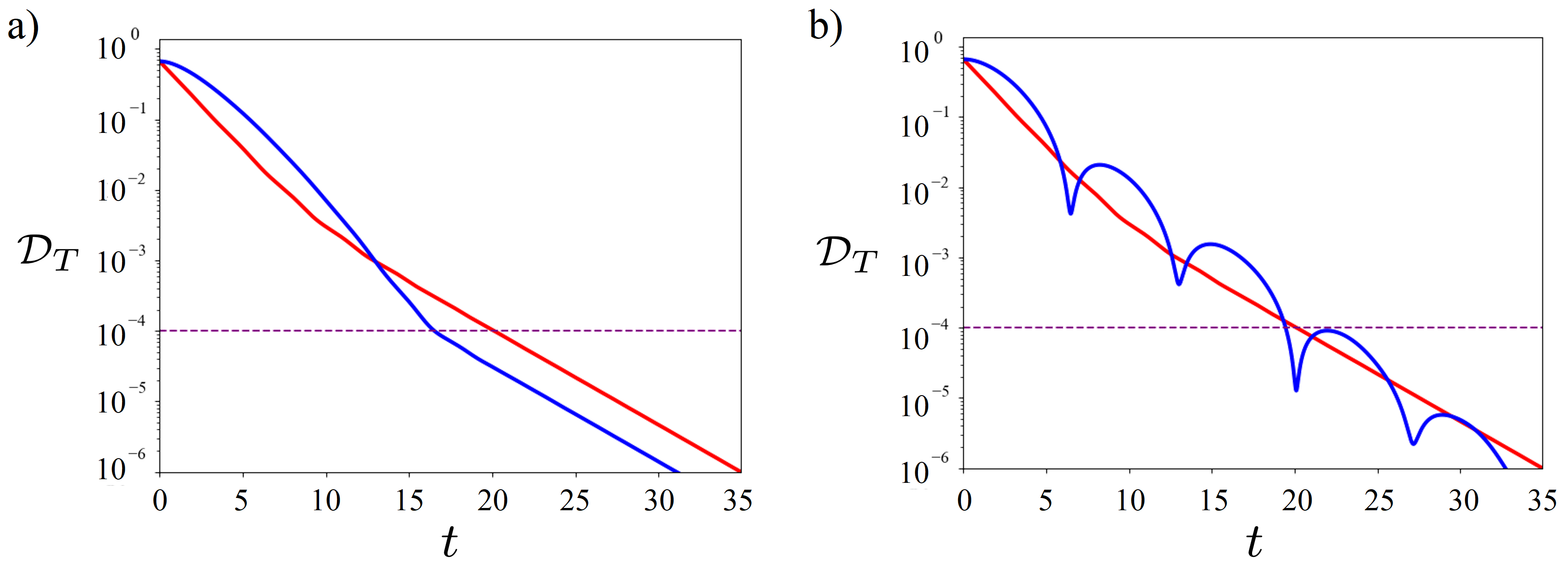}
    \caption{ Comparison between direct and continuous Pontus-Mpemba protocols for open Markovian two-state systems.
    \textbf{(a)}  Trace distance $\mathcal{D}_T \left( \rho(t), \rho_\mathrm{F} \right)$ vs time $t$. For both protocols, the initial state S is the steady state \eqref{eq:steady-state} corresponding to $\mathbf{h}_\mathrm{S}=(0.183,0.183,-0.966)$ and
    $(\gamma_{+},\gamma_-,\gamma_z)_{\rm S}=(0.5,0.1,0),$ while the final state F 
    is determined by $\mathbf{h}_\mathrm{F}=\mathbf{h}_\mathrm{S}$ and $(\gamma_{+},\gamma_-,\gamma_z)_{\rm F}=(0.1,0.5,0)$. 
    The trace distance for the direct protocol ${\rm S}\rightarrow{\rm F}$ is shown as red curve; the dashed line is the cutoff $\epsilon$. The blue curve shows the trace distance between S and F for the time-dependent rates in Eq.~\eqref{eq:gamma_time-dependent} with $\kappa=0.6$ and $\omega=0.2$. 
    Both protocols reach the same final state F for $t\to \infty$.  Clearly, the continuous Pontus-Mpemba effect is realized, and a crossing of the trace distance curves takes place at $t\approx 13$.
    \textbf{(b)}  Same as (a) but for $\kappa=0.4$ and $\omega=0.45$. The inconclusive regime is observed.
    }
    \label{fig:fig3}
\end{figure}

Below we compare the direct (sudden-quench quantum Mpemba) protocol with the continuous Pontus-Mpemba protocol. For this purpose, we define a ``gain function'' as 
\begin{equation}\label{gain_function}
    G(\kappa,\omega)=\frac{\tau_\mathrm{dir}}{\tau_\mathrm{cPM}(\kappa,\omega)} -1,
\end{equation}
where $\tau_\mathrm{dir}$ is the relaxation time for the direct protocol and $\tau_\mathrm{cPM}$ the relaxation time for the continuous Pontus-Mpemba protocol with given $\kappa$ and $\omega$. The gain function vanishes if both approaches predict the same relaxation time, while $G<0$ if the direct protocol is faster.  If the continuous Pontus-Mpemba effect brings an advantage, we have $G>0$. If ${\cal F}>0$ holds in addition, we have a non-Markovian continuous Pontus-Mpemba effect.
For $\kappa \rightarrow 0$, the time evolution is quasi-static, and the evolution speed cannot exceed the speed at which the rates (and thus the attractor) evolve. In this limit, we have $\tau_\mathrm{cPM}(\kappa \rightarrow 0,\omega) \rightarrow \infty$ and therefore $G\rightarrow -1$, i.e., the quasi-static protocol is always slower than the direct protocol. On the other hand, for the sudden-quench limit $\kappa \rightarrow \infty$, we have by definition $\tau_\mathrm{cPM} \rightarrow \tau_\mathrm{dir}$ and thus $G= 0$.
By tuning $\kappa$ in between these two limits, we explore  a crossover regime   which may allow for transient nonequilibrium effects causing $G>0$, resulting in a speed-up with respect to the sudden-quench protocol. In Figs.~\ref{fig:fig2} and \ref{fig:fig3}, we present representative examples highlighting the emergence of such nontrivial transient behavior and possible speed-ups. 

In Fig.~\ref{fig:fig2}(a), we show $\mathcal{D}_T \left( \rho(t), \rho_\mathrm{F} \right)$ as a function of time for the direct protocol (red curve) and two non-autonomous cases. Under the direct protocol, the trace distance reaches the cutoff value $\epsilon$ at the time $\tau \approx 160$. The blue curve, obtained for $\kappa=0.2$, exhibits a continuous Pontus-Mpemba effect with $\tau \approx 60$, resulting in the gain $G\approx 1.66$. 
By further decreasing $\kappa$, one eventually destroys the continuous Pontus-Mpemba effect, as shown by the green curve for $\kappa=0.035$, where we obtain $\tau \approx 200$ and thus $G<0$. 

Thanks to the explicit time dependence of the rates it becomes possible to dynamically reshape the relaxation trajectory in state space. For the direct protocol, as shown, e.g., in Fig.~\ref{fig:fig2}(b),  the Bloch vector $\mathbf{r}(t)$ typically follows a damped spiral-type path toward the steady state dictated by the dissipative structure of the Lindbladian and by the field. By appropriately tuning the time dependence of the rates, however, one can effectively ``cut across'' this spiral motion, thereby shortening the geometric path
taken by ${\bf r}(t)$ inside the Bloch ball, see Figs.~\ref{fig:fig2}(c,d). The system is then steered along a shorter trajectory toward the target state, with the possibility of reducing the overall relaxation time.
This mechanism provides a second design principle for speed-up. The first principle (discussed above) relies on driving the system into state space regions where the velocity field has a large amplitude. The second principle instead exploits the possibility of reshaping the trajectory itself, thereby reducing its length by dynamically redirecting the flow in state space. The combined action of these two effects, i.e., accessing high-velocity regions and shortening the relaxation trajectory, can allow the system to reach the target state faster than under the direct protocol. We refer to these trajectories as ``dynamical shortcuts'' since their presence cannot be forecasted by just looking at the   velocity field for the final state only. Instead, such shortcuts depend on the history of the system and are thus dynamically generated.
However, this advantage is not guaranteed to exist for arbitrary driving parameters $(\kappa,\omega)$. In particular, if $\kappa$ is too small, the evolution approaches the quasi-static regime, where the system adiabatically follows the instantaneous attractor. In this limit, the geometric shortening mechanism becomes ineffective and 
the time gained by this mechanism will progressively diminish (leading to $G\rightarrow -1$) even though the path length is reduced.    Therefore a speed-up emerges only in an intermediate-$\kappa$ regime.

Before investigating whether this effect requires fine tuning of the parameters $(\kappa,\omega)$, which would limit its experimental relevance, or whether it instead persists over a broad parameter region, we present in Fig.~\ref{fig:fig3} two additional examples for the time evolution of the trace distance. Similar to the two-step Pontus-Mpemba protocol, also the continuous Pontus-Mpemba effect can be realized with or without  a crossing point between the direct and the non-autonomous trace distance curves. The example in Fig.~\ref{fig:fig2}(a), comparing the red and blue curves, shows a case without  crossing point. In Fig.~\ref{fig:fig3}(a), we show a case exhibiting a crossing point. 
In general, since $\mathcal{D}_T \left( \rho_{\rm cPM}(t), \rho_\mathrm{F} \right)$ is not necessarily a monotonic function of time for the non-autonomous Lindblad equation \eqref{eq:Lindblad_ME}, we may distinguish two general cases by comparing the initial decay rates of the trace distance between the time-dependent (continuous Pontus-Mpemba) and the direct (${\rm S}\rightarrow{\rm F})$ protocol. In particular, if
\begin{equation}
    \frac{ d }{d t} \mathcal{D}_T \left( \rho_{\rm cPM}(0), \rho_\mathrm{F} \right) < \frac{ d}{d t} \mathcal{D}_T \left( \rho_{\rm SF}(0), \rho_\mathrm{F} \right),
\end{equation}
such that the trace distance initially decreases faster for the continuous Pontus-Mpemba protocol, the engineered protocol exhibits an initial acceleration toward the target state. The continuous Pontus-Mpemba effect then could take place if the two trace distance curves cross an even number of times. Conversely, if
\begin{equation}
    \frac{d}{d t} \mathcal{D}_T \left( \rho_{\rm cPM}(0), \rho_\mathrm{F} \right) > \frac{d}{dt} \mathcal{D}_T \left( \rho_{\rm SF}(0), \rho_\mathrm{F} \right),
\end{equation}
the relaxation is initially faster for the direct-quench case. In this case, an odd number of crossing points is required to have the continuous Pontus-Mpemba effect.
Similarly to the notation adopted for the two-step Pontus-Mpemba effect, we  refer to those two cases as \textit{weak} and \textit{strong} continuous Pontus-Mpemba effects, respectively.

We next observe that oscillations in the trace distance curves are suppressed on time scales of order $\tilde t\approx\kappa^{-1}$  since the rate modulation  
progressively attenuates toward  $\gamma_{\lambda,\mathrm{F}}$. However, if during this transient window, the trace distance reaches the prescribed cutoff $\epsilon$ before the oscillations are damped out, the comparison between both protocols becomes ambiguous. In this situation, the crossing of the threshold may occur during an oscillatory excursion, making it impossible to unambiguously determine which protocol provides a genuinely faster relaxation.
We refer to this scenario as the inconclusive regime, where the presence of transient oscillations prevents a clear identification of a speed-up  with respect to the 
direct-quench protocol. An example for this case is shown in Fig.~\ref{fig:fig3}(b).

\begin{figure}
    \centering
    \includegraphics[width= 0.9 \textwidth]{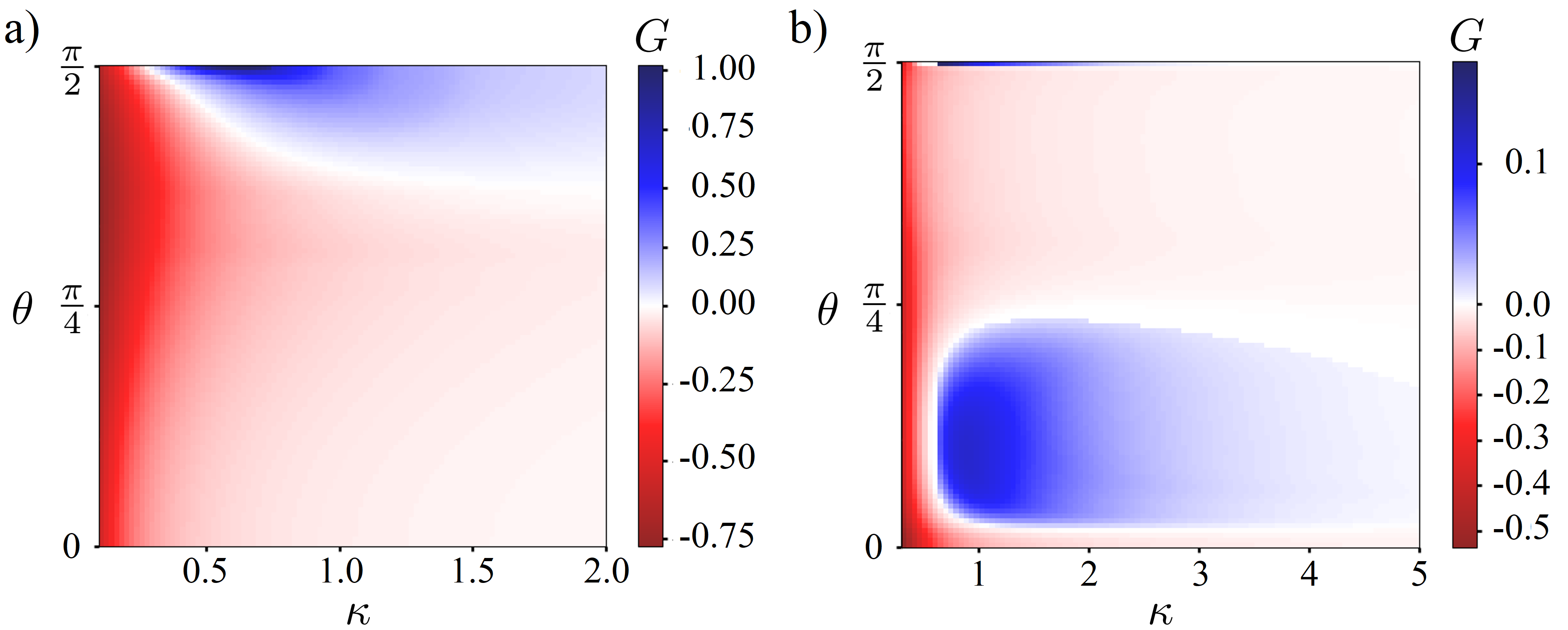}
    \caption{ Color-scale plots of the gain function $G(\kappa,\omega)$ in Eq.~\eqref{gain_function} for $\omega=0$, where $G>0$ indicates a speed-up under to the continuous Pontus-Mpemba protocol with respect to the direct sudden-quench protocol. Note that the dynamics is then always Markovian.
    \textbf{(a)}    $G$ in the $\kappa$--$\theta$ plane, where $\theta$ is the angle between $\mathbf{h}_\mathrm{F}=\mathbf{h}_\mathrm{S}=(\sin{\theta},0,\cos{\theta})$ and the $z$-axis. The initial state S is determined by $(\gamma_{+},\gamma_-,\gamma_z)_\mathrm{S}=(0.75, 0.75, 0.75)$,   and the final
    state F corresponds to  $(\gamma_{+},\gamma_-,\gamma_z)_\mathrm{F}=(0.05, 0.1, 0.15).$  
    \textbf{(b)}  Same as (a) but for $(\gamma_{+},\gamma_-,\gamma_z)_\mathrm{S}=(0.5,0.1,0)$ and
    $(\gamma_{+},\gamma_-,\gamma_z)_\mathrm{F}=(0.1,0.5,0)$.
    }
    \label{fig:fig4}
\end{figure}

In Fig.~\ref{fig:fig4}, we show color-scale plots for the dependence of the gain function $G(\kappa,\omega)$ in Eq.~\eqref{gain_function} on $\kappa$ and $\theta$ for a pure decay of the time-dependent rates, i.e., for $\omega=0$. Here the angle $\theta$ refers to the angle between the field $\mathbf{h}$ (which is kept constant before and after the quench) 
and the $z$-axis. As expected, in the quasi-static limit $\kappa \rightarrow 0$, the continuous Pontus-Mpemba protocol gives $G\rightarrow -1$ (red region). 
In the opposite limit $\kappa \rightarrow \infty$, the protocol reduces to the direct sudden-quench case, yielding $G\rightarrow 0$ (white region).
However, for intermediate values of $\kappa$, extended regions with $G>0$ emerge (blue), indicating a genuine speed-up that does not require a fine tuning of parameters. 
In Fig.~\ref{fig:fig4}(a), a single broad $G>0$ region appears if $\mathbf{h}$ is approximately perpendicular to the $z$-axis. For the parameters chosen in
Fig.~\ref{fig:fig4}(b), the structure is richer. Apart from a narrow region around the perpendicular configuration, an additional wider region of positive gain develops for smaller values of $\theta$.
Importantly, varying the field direction modifies only the unitary dynamics, leaving the dissipative channels unchanged. This separation provides a particularly clean experimental handle: by tuning the field orientation, one can probe the relaxation process without altering the structure of the environment-induced dissipation. As a consequence, the emergence of regions with positive gain under such variations offers a direct and experimentally accessible signature for continuous Pontus-Mpemba effects.
However, it is important to stress that varying the angle $\theta$ non-trivially shifts both the initial steady state S and the final target state F. As a consequence, changes in $\theta$ simultaneously affect the geometry of the relaxation trajectory $\mathbf{r}(t)$ and the relative positioning of the attractors in state space.

\begin{figure}
    \centering
    \includegraphics[width= 0.9 \textwidth]{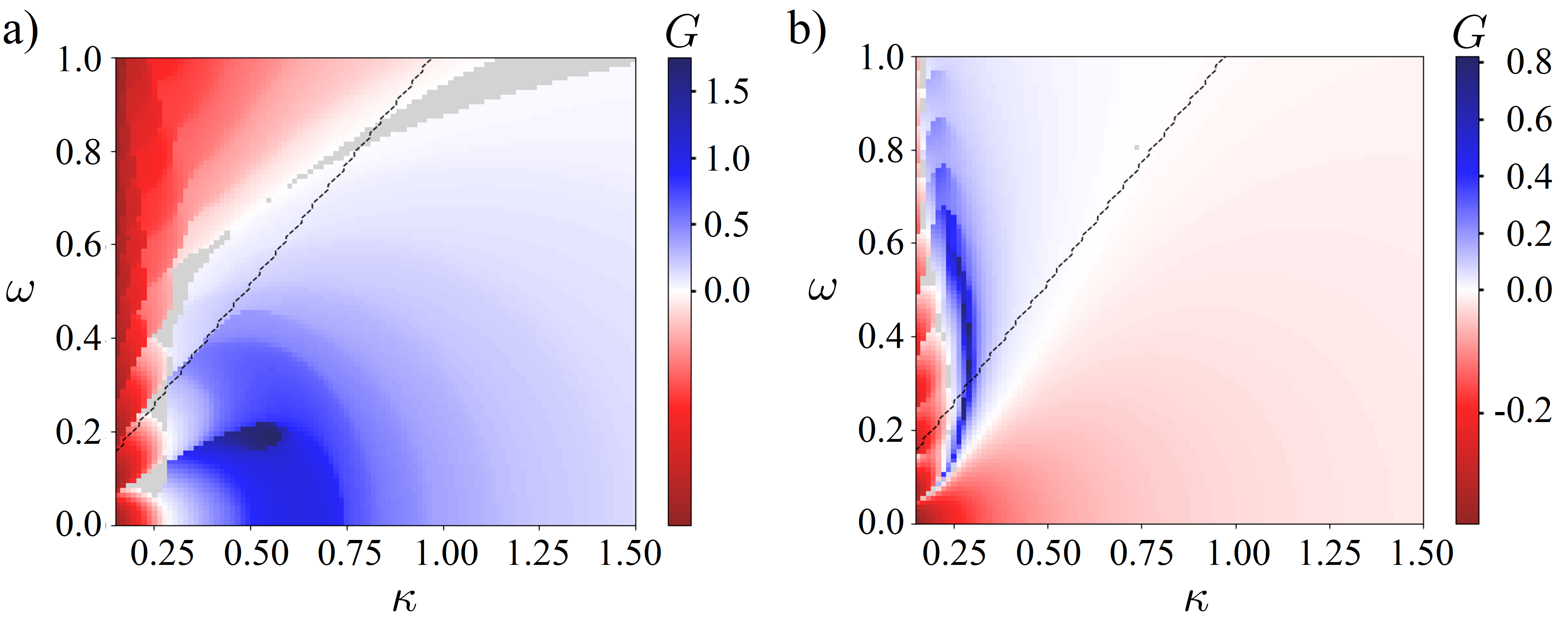}
    \caption{Color-scale plots for the gain function $G(\kappa,\omega)$ in Eq.~\eqref{gain_function} for $\omega>0$ in Eq.~\eqref{eq:gamma_time-dependent}.
    Dashed lines separate Markovian (below) and non-Markovian (above) regimes, see Eq.~\eqref{nMB}. Gray areas correspond to the inconclusive regime.
    \textbf{(a)}  $G$ in the $\kappa$--$\omega$ plane for $\mathbf{h}=\mathbf{h}_\mathrm{F}=\mathbf{h}_\mathrm{S}=(1,0,0)$ perpendicular to the $z$-axis. 
    All other parameters are as in Fig.~\ref{fig:fig4}(a).
    \textbf{(b)} Same as (a) but for $\mathbf{h}=(0,0,1)$ parallel to the $z$-axis.  (Again,  other parameters are as in Fig.~\ref{fig:fig4}(a).)
    }
    \label{fig:fig5}
\end{figure} 

Although  at fixed $\theta$,  no critical value of $\kappa$ exists above which the continuous Pontus-Mpemba effect is automatically triggered, 
the effect can nevertheless be induced by changing the oscillation frequency $\omega$ in Eq.~\eqref{eq:gamma_time-dependent}. In Fig.~\ref{fig:fig5}, we show the gain $G$ as
color-scale plot in the $\kappa$--$\omega$ plane for the parameters in Fig.~\ref{fig:fig4}(a).  In Fig.~\ref{fig:fig5}(a) and (b), we  consider fields $\mathbf{h}$ which are  perpendicular ($\theta=\frac{\pi}{2}$) and parallel ($\theta=0$) to the $z$-axis, respectively.
In both cases, the regions of positive gain are robust against variations of $\omega$, persisting over extended frequency intervals. Moreover, additional gain regions, which are absent for $\omega=0$, can emerge for $\omega>0$. This demonstrates that the oscillatory component of the modulation provides an additional control knob which is capable of activating the continuous Pontus-Mpemba effect even in parameter regimes where it would otherwise be absent.

Next we note that by increasing $\omega$, the dynamics can enter a non-Markovian regime, as signaled by the breakdown of CP-divisibility and a non-zero value of the measure of non-Markovianity $\mathcal{F}$. The corresponding transition line has been specified in Eq.~\eqref{nMB}. Although we do not observe a direct systematic correlation between the emergence of the continuous Pontus-Mpemba effect and the onset of non-Markovianity, this regime is of conceptual and practical interest.
First, in the non-Markovian regime, the instantaneous generator may correspond to a transient attractor that formally lies outside the Bloch ball. While such an attractor is not a physical state per se, this additional flexibility can enhance the ability to reshape trajectories in state space, thereby increasing the range of protocols capable of exhibiting a continuous Pontus-Mpemba effect. In this sense, non-Markovianity may broaden the controllable landscape of relaxation paths.
Second, from an experimental perspective, the absence of a strict link between the continuous Pontus-Mpemba effect and (non-)Markovianity 
suggests that this effect is observable in a wider class of open quantum systems, including those not accurately described by time-local Markovian Lindblad equations. This enlarges the set of candidate platforms where such relaxation speed-ups might be realized and detected.

\begin{figure}
    \centering
    \includegraphics[width= 0.9 \textwidth]{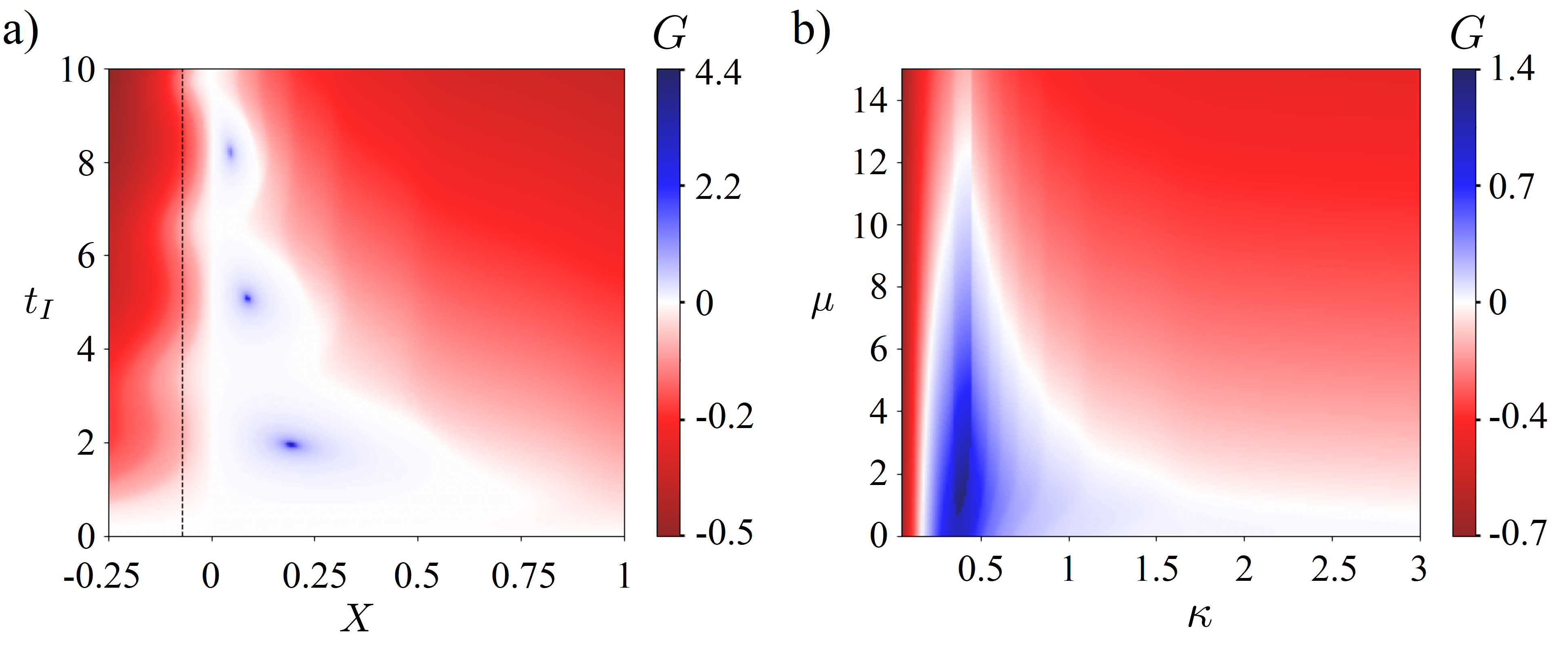}
    \caption{\new{Color-scale plots of the gain function $G$ for two different families of time-dependent rates beyond Eq.~\eqref{eq:gamma_time-dependent}. In both panels, we consider a constant Hamiltonian with $\mathbf{h}=\mathbf{h}_\mathrm{F}=\mathbf{h}_\mathrm{S}=(1,0,0)$. The rates $(\gamma_{+},\gamma_-,\gamma_z)_\mathrm{S}=(0.75, 0.75, 0.75)$ and $(\gamma_{+},\gamma_-,\gamma_z)_\mathrm{F}=(0.05, 0.1, 0.15)$ determine the initial and final states, respectively, as in Fig.~\ref{fig:fig4}(a).
    \textbf{(a)} Two-step Pontus-Mpemba protocol \cite{Nava2025}, with the time-dependent rates in Eq.~\eqref{eq:gamma_time-dependent_2} parametrized by $X$ and $t_{\rm I}$.   We show the gain function $G$ in the $X$--$t_{\rm I}$ plane.  For $X=0$, the standard one-step quantum Mpemba protocol follows, resulting in $G=0$. 
    The dashed vertical line separates Markovian (right) and non-Markovian (left) regimes, see Eq.~\eqref{nMB_2}.  
    \textbf{(b)} $G$ in the $\kappa$--$\mu$ plane for the shifted hyperbolic tangent interpolation family of time-dependent rates in Eq.~\eqref{eq:gamma_time-dependent_3} parameterized by $\kappa$ and $\mu$. Here the dynamics is always Markovian.
    }}
    \label{fig:fig6}
\end{figure}

\new{Finally, in order to assess the robustness of the observed relaxation speed-up with respect to changing the time dependence of the rates, we consider additional families of time-dependent protocols beyond Eq.~\eqref{eq:gamma_time-dependent}. In particular, we study (i) the two-step protocol \cite{Nava2025} and (ii) a shifted hyperbolic tangent interpolation.
The corresponding results for the gain function are shown in Fig.~\ref{fig:fig6}.}
\new{For the two-step protocol, we assume that the time-dependent rates are given by 
\begin{equation}
    \gamma_{\lambda}(t>0) = \gamma_{\lambda,\mathrm{A}}\: \Theta(t_\mathrm{I} - t) + \gamma_{\lambda,\mathrm{F}} \: \Theta(t - t_\mathrm{I}),\quad \gamma_{\lambda,\mathrm{A}}
    =\gamma_{\lambda,\mathrm{F}} + X \left( \gamma_{\lambda,\mathrm{S}} - \gamma_{\lambda,\mathrm{F}}\right),
    \label{eq:gamma_time-dependent_2}
\end{equation}
where $t_\mathrm{I}$ sets the duration of the intermediate stage before the final quench.  The auxiliary state A is determined by the rates $\gamma_{\lambda,\mathrm{A}}$, which are here (for simplicity) parametrized by a single parameter $X$.  Clearly, for $X=0$, we recover the direct (one-step) quantum Mpemba protocol at $t=0^+$. 
For the case $\gamma_{\lambda,\mathrm{S}}>\gamma_{\lambda,\mathrm{F}}$ corresponding to Fig.~\ref{fig:fig6}(a), we have $\gamma_{\lambda,\mathrm{F}}<\gamma_{\lambda,\mathrm{A}}<\gamma_{\lambda,\mathrm{S}}$ for $0<X<1$, while for $X<0$, one finds $\gamma_{\lambda,\mathrm{A}}<\gamma_{\lambda,\mathrm{F}}$.  From Eq.~\eqref{nMB}, we then see that
for $X<-\gamma_{\lambda,\mathrm{F}}/(\gamma_{\lambda,\mathrm{S}}-\gamma_{\lambda,\mathrm{F}})$, one enters a non-Markovian regime where the non-Markovianity measure for each channel is given by
\begin{equation}
    \mathcal{F}_\lambda = t_\mathrm{I} \: \gamma_{\lambda,\mathrm{A}}.
    \label{nMB_2}
\end{equation}
The separation between the Markovian and non-Markovian regimes is indicated by a vertical dashed line in Fig.~\ref{fig:fig6}(a).
A comparison between the two-step Pontus-Mpemba and the direct quantum Mpemba protocol is now performed in terms of the gain function \eqref{gain_function} by replacing $\tau_{\rm cPM}(\kappa,\omega)$ with the relaxation time $\tau_{\rm 2sPM}(X,t_\mathrm{I})$ obtained under the two-step protocol. In Fig.~\ref{fig:fig6}(a), we show a color-scale plot for the resulting gain function $G(X,t_\mathrm{I})$. An extended region with $G>0$ emerges at intermediate values of the control parameters ($X,t_{\rm I})$, which is dominated by sharply peaked islands centered around fine-tuned 
parameter values. On such peaks, $G$ can reach higher values than 
for the continuous Pontus-Mpemba case in Fig.~\ref{fig:fig5}. However, when moving away from the optimal parameter values realizing a peak, $G$ quickly drops down to tiny values. For this specific example, we conclude that the two-step Pontus-Mpemba protocol can be effective in speeding up the relaxation process.  However, compared to the continuous Pontus-Mpemba protocol, it lacks robustness in the sense that it requires a precise fine tuning of parameters.  Next we consider a shifted hyperbolic tangent interpolation for the time dependence of the rates,
\begin{equation}
    \gamma_{\lambda}(t>0) = \gamma_{\lambda,\mathrm{F}} 
    + \left(\gamma_{\lambda,\mathrm{S}} - \gamma_{\lambda,\mathrm{F}}\right) 
    \frac{1 - \tanh[\kappa(t - \mu)]}{1 + \tanh(\kappa\mu)},
    \label{eq:gamma_time-dependent_3}
\end{equation}
where the rate $\kappa$ again corresponds to the inverse time scale governing the crossover between the initial and final states. In particular, large values of $\kappa$ imply a sharp quench-like behavior, while small $\kappa$ amounts to a slow quasi-static time evolution. 
The parameter $\mu$ determines the time around which the transition between the initial and the final state is centered. For Eq.~\eqref{eq:gamma_time-dependent_3}, non-Markovian effect are not possible since the time-dependent rates are always confined between their initial and final values. 
In Fig.~\ref{fig:fig6}(b), we show a color-scale plot for the resulting
gain function $G(\kappa,\mu)$, defined as in Eq.~\eqref{gain_function} by replacing $\tau_{\rm cPM}(\kappa,\omega)$ with the relaxation time $\tau_{\rm cPM}(\kappa,\mu)$. Similar to what we reported for  the time-dependent rates in Eq.~\eqref{eq:gamma_time-dependent}, see, e.g., Fig.~\ref{fig:fig5}, we obtain an extended region of positive gain, $G>0$, which does not require any fine tuning of control parameters. 
In conclusion, we find that the relaxation speed-up under continuous Pontus-Mpemba protocols is found for qualitatively different families governing the time dependence of the rates. 
}

\section{Discussion and Conclusions}
\label{conclusions}

In this work, we have investigated the possibility of accelerating quantum state preparation in an open quantum system through a non-autonomous Lindblad equation by modulating the generators of the unitary and dissipative dynamics. Moving beyond the quasi-static regime, where the evolution adiabatically follows a slowly varying attractor, and the sudden-quench limit, where parameters are abruptly switched, we have shown that in the intermediate crossover regime, the interplay between coherent evolution and dissipation gives rise to a nontrivial dynamical reshaping of the relaxation trajectories.
Within this framework, we have analyzed a continuous version of the Pontus-Mpemba protocol, where an infinite sequence of infinitesimal quenches generates a continuous driving with time-dependent Hamiltonian as well as dissipation parameters.  For the example of an open two-level quantum system, we have identified parameter regions where a genuine speed-up occurs. In particular, we showed the existence of extended regions characterized by a shorter relaxation time as compared to both the quasi-static and sudden-quench protocols.  These regions are robust in the sense that they represent broad regions with stable acceleration gain in the phase diagram, such that fine tuning of parameters is not necessary. 

Geometrically, the mechanism underlying the speed-up can be interpreted in two complementary ways. First, one can steer the system toward regions in state space that are characterized by faster relaxation rates, thereby reducing the projection of the initial state onto eigenstates corresponding to slow modes. Second, the explicit time dependence of the rates allows one to reshape the state trajectory itself. In this way, one effectively shortens the path toward the target state by ``cutting across'' the relaxation path followed by the system  trajectory under a sudden-quench protocol. The combined action of these
mechanisms can yield a net reduction in the time required to reach a prescribed accuracy for the target state. \new{In this paper, we have focused on cases where the time dependence is introduced via the dissipation rates of the Lindblad generator. However, the intermediate state can also by reached via time-dependent Hamiltonians, thus realizing unitary control protocols. In this way, it is possible to dynamically engineer the instantaneous unitary transformations proposed, 
e.g., in Refs.~\cite{Carollo_2021,Kochsiek2022,Moroder2024} in order to achieve an exponential speed-up.}

\new{While we have studied a two-level system, we expect that that our general conclusions are valid also 
for open quantum many-body systems.
In   systems with many degrees of freedom, the spectrum of the Lindbladian typically becomes dense, potentially leading to a manifold of slow relaxation modes with comparable decay rates such that suppressing only the projection onto the slowest mode may not produce a significant speed-up.  In order to address this issue, we emphasize that the continuous Pontus-Mpemba effect relies on two main ingredients, namely (i) the identification of fast relaxation regions as illustrated in Fig.~\ref{fig:fig1}, and (ii) the dynamical trajectory reshaping mechanism shown in Fig.~\ref{fig:fig2}.  The denseness of the spectrum of the Lindbladian for a many-body system may influence the first mechanism, while the second mechanism should not be significantly affected.  In addition, even for the standard quantum Mpemba effect, robustness is achieved for a many-body system by suppressing the projection onto the \emph{manifold} of slow modes (and not just the single slowest mode) while searching
for velocity-field regions with high relaxation speed.  Indeed, quantum Mpemba effects have been numerically found in systems with many degrees of freedom. For example, Ref.~\cite{Gal_2020} has studied a two-dimensional Ising model on an $L\times L$ square lattice with $L=70$, while in Ref.~\cite{Nava_2025_2},   both Mpemba and Pontus-Mpemba effects have been reported for the mean-field Gross-Neveu model.}

We have also clarified the role of non-autonomous and potentially non-Markovian dynamics. While the breakdown of CPT-divisibility is not a necessary condition for the emergence of the continuous Pontus-Mpemba effect, entering a non-Markovian regime expands the range of accessible protocols by increasing the flexibility concerning the instantaneous attractor. \new{The lack of a direct correlation between non-Markovian features (i.e., negative rates) and relaxation speed-up indicates that non-Markovianity is neither a necessary nor a sufficient condition for the continuous Pontus-Mpemba effect. Rather, the speed-up originates from the interplay between the time-dependent attractor and the geometry of the dynamical trajectories.} Our results highlight that relaxation times can be engineered through a controlled time-dependent modulation of dissipation rates, representing a complementary approach to state preparation when compared to QOC protocols relying on engineered unitary dynamics. 

From an experimental perspective, the protocols considered here are compatible with platforms where dissipation rates and effective Hamiltonian parameters can be tuned dynamically. Examples include superconducting qubits with engineered reservoirs \cite{Shankar2013,Sevriuk_2019,Li2025}, trapped ions with controlled dissipation channels \cite{So2024,So2026}, cold atoms in optical lattices coupled to tailored environments \cite{McKay2008,Viebahn_2021}, and spin systems \cite{Kampermann_2006}.

Our paper naturally opens several interesting questions. A first direction for future research concerns robustness, i.e., to what extent will the parameter regions showing acceleration persist for imperfect control over the dissipation channels? A second issue is about optimality, i.e., how should one design the time-dependence of the rates, the Hamiltonian, and, eventually, of the jump operators, in order to achieve the highest possible speed-up in a given system? 
A third question concerns scalability and how these mechanisms generalize to open systems with many degrees of freedom  \cite{Kshetrimayum2017,Nava_2023,Alpomishev_2024,Nava_2024_2,Morales_2024,Morales_2025}, possibly exhibiting topological order \cite{Diehl2011,Gneiting_2022,Cinnirella_2024,Cinnirella_2025,Cinnirella_2026,Pokart2026} and/or featuring strong particle interactions \cite{Gorshkov_2013,Nava_2022,Boneberg_2023,Berger_2025,Sieberer_2025}. We leave these questions for future research. 

\new{We emphasize that our work studies a restricted family of time-dependent rates, where the time dependence of the rates between their initial and final values
is determined by two parameters ($\omega$ and $\kappa$). 
It is natural to ask whether truly optimal control strategies could further enhance the reported relaxation speed-up. In principle, this problem can be formulated within the framework of QOC, where the dissipative rates and Hamiltonian parameters are treated as control functions and the objective is to minimize the time required to reach the target state, see, e.g., Ref.~\cite{Gal_2020} for the classical problem of a four-state system under the effect of a time-dependent bath temperature. Recent advances in QOC have shown that machine-learning techniques, such as reinforcement learning and neural-network-based optimization, can be used to discover nontrivial time-dependent control protocols for low-dimensional systems \cite{Wauters_2020,Metz2023,Li2025_3}. However, our setting presents several challenges to such an endeavour. In particular, the requirement of physicality of the dynamics imposes nontrivial constraints, such as complete positivity or positivity conditions that are generally nonlocal in time. To avoid those complications, we have here pursued a complementary approach based on a physically motivated family of protocols, leaving the exploration of systematic (say, machine-learning-based) optimization protocols to future research.}

In conclusion, the continuous Pontus-Mpemba protocol provides a paradigm in which the topics of nonequilibrium dynamics, geometry of state space, and controlled dissipation meet. It applies to both classical and quantum systems.  Within the quantum domain, this framework can allow  for accelerated state preparation protocols, including options for fast
real-time manipulation of open quantum systems.

\ack{We thank D. Giuliano and I. Gornyi for discussions.}
\funding{M.P. is funded through DM 118/2023 - Inv. 4.1, project ``Light-matter interactions in topological semimetals,'' CUP E14D23001640006, PNRR. Further, R.E. and A.N. acknowledge funding by the Deutsche Forschungsgemeinschaft (DFG, German Research Foundation) under Projektnummer 277101999 - TRR 183 (project C01), under Project No. EG 96/13-1, and under Germany's Excellence Strategy - Cluster of Excellence Matter and Light for Quantum Computing (ML4Q) EXC 2004/2 - 390534769.}

\roles{All authors contributed equally to the research and the preparation of the manuscript.}  

\data{All data supporting the findings of this paper are available on Zenodo \cite{zenodo}.}

\bibliographystyle{unsrturl}
\bibliography{references}

\end{document}